\documentclass[aps,pra,twocolumn,amsmath,amssymb,nofootinbib,superscriptaddress]{revtex4-1}

\usepackage{times}
\usepackage[pdftex]{graphicx}
\usepackage{dcolumn}
\usepackage{bm}
\usepackage{amsmath}
\usepackage{indentfirst}
\usepackage{float}
\usepackage[colorlinks]{hyperref}
\usepackage[dvipsnames]{xcolor}

\usepackage{xcolor}

\usepackage[normalem]{ulem}

\newcommand{\sgz}{Z^{\otimes N}}

\newcommand{\pcircuit}{\mathcal{C}(\vec{\bold{\theta})}}
\newcommand{\Ry}{R_y}

\newcommand{\gcc}[1]{{\color{black}{#1}}}
\newcommand{\hhl}[1]{{\color{black}{#1}}}

\newcommand{\QCCNN}{{\rm QCCNN}}

\newcommand{\istd}{Information Systems Technology and Design, Singapore University of Technology and Design, 8 Somapah Road, 487372 Singapore}

\newcommand{\idc}{SUTD-MIT International Design Center, Singapore University of Technology and Design, 8 Somapah Road, 487372 Singapore}

\newcommand{\epd}{Engineering Product Develoment, Singapore University of Technology and Design, 8 Somapah Road, 487372 Singapore}

\begin{document}

\title{Hybrid Quantum-Classical Convolutional Neural Networks}

\author{Junhua Liu}
\affiliation{\istd}
\affiliation{\idc}

\author{Kwan Hui Lim}
\affiliation{\istd}

\author{Kristin L. Wood}
\affiliation{\idc}
\affiliation{\epd}



\author{Wei Huang}
\affiliation{Guangxi Key Laboratory of Optoelectronic Information Processing, Guilin University of Electronic Technology, Guilin 541004, China}

\author{Chu Guo}
\email{guochu604b@gmail.com}
\affiliation{Key Laboratory of Low-Dimensional Quantum Structures and Quantum Control of Ministry of Education, Department of Physics and Synergetic Innovation Center for Quantum Effects and Applications, Hunan Normal University, Changsha 410081, China}

\author{He-Liang Huang}
\email{quanhhl@ustc.edu.cn}

\affiliation{Hefei National Laboratory for Physical Sciences at Microscale and Department of Modern Physics,\\
University of Science and Technology of China, Hefei, Anhui 230026, China}
\affiliation{Shanghai Branch, CAS Centre for Excellence and Synergetic Innovation Centre in Quantum Information and Quantum Physics,\\
University of Science and Technology of China, Hefei, Anhui 201315, China}
\affiliation{Henan Key Laboratory of Quantum Information and Cryptography, Zhengzhou, Henan 450000, China}

\date{\today}


\begin{abstract}
Deep learning has been shown to be able to recognize data patterns better than humans in specific circumstances or contexts. In parallel, quantum computing has demonstrated to be able to output complex wave functions with a few number of gate operations, which could generate distributions that are hard for a classical computer to produce. Here we propose a hybrid quantum-classical convolutional neural network (QCCNN), inspired by convolutional neural networks (CNNs) but adapted to quantum computing to enhance the feature mapping process.
QCCNN is friendly to currently noisy intermediate-scale quantum computers, in terms of both number of qubits as well as circuit's depths, while retaining important features of classical CNN, such as nonlinearity and scalability. We also present a framework to automatically compute the gradients of hybrid quantum-classical loss functions which could be directly applied to other hybrid quantum-classical algorithms. We demonstrate the potential of this architecture by applying it to a Tetris dataset, and show that QCCNN can accomplish classification tasks with learning accuracy surpassing that of classical CNN.
\end{abstract}

\maketitle

\section{Introduction}\label{sec:intro}

With the rapid progress in quantum computing hardware, we are entering the era of developing quantum software to perform useful computational tasks using the noisy intermediate-scale quantum (NISQ) computers~\cite{preskill2018quantum,huang2020superconducting}. It has been shown in 2019 that the current $53$-qubit quantum computer could already solve random quantum circuit sampling problems more efficiently than the best supercomputers in the world~\cite{arute2019quantum}. Theoretically, this success originates from the fact that a quantum computer could output wave functions with polynomial number of quantum gate operations, which could nevertheless generate statistical distributions that are very hard for a classical computer to produce~\cite{harrow2017quantum,GuoWu2019,guo2021verifying}. \gcc{Recently it is reported that the $53$-qubit $20$-depth random quantum circuit sampling problem has been successfully simulated classically~\cite{PanZhang2021}, however, simulating larger-scale quantum circuits is still a huge challenge.} If a quantum computer could easily produce complex distributions, it is also natural to postulate that it is able to learn patterns from certain data distributions which could be very difficult for classical computers~\cite{biamonte2017quantum}.
Quantum machine learning (QML) attempts to utilize this power of quantum computer to achieve computational speedups or better performance for machine learning tasks~\cite{biamonte2017quantum, lloyd2014quantum, lloyd2016quantum, rebentrost2014quantum, huang2018demonstration,huang2017homomorphic,ding2019quantum}, and parameterized quantum circuits (PQCs) offer a promising path for quantum machine learning in the NISQ era~\cite{mcclean2016theory,benedetti2019parameterized,LiuWu2020}. \hhl{Compared to traditional quantum algorithms~\cite{Shor1994, grover1996fast, long2001grover,huang2018demonstration} such as Shor's algorithm~\cite{Shor1994,lu2007demonstration,huang2017experimental}, PQC-based quantum machine learning algorithms are naturally robust to noise~\cite{gentini2020noise} and only require shallow quantum circuits, which \gcc{is} highly desirable for near-term noisy intermediate scale quantum devices. Compared to classical neural networks, PQC-based quantum machine learning algorithms have two main potential advantages, stronger expressive power~\cite{killoran2019continuous,du2020expressive,coyle2020born,tangpanitanon2020expressibility,schuld2019quantum,SarmaDuan2019} and stronger computing power~\cite{lloyd2018quantum}, originating from the superposition principle of quantum mechanics. Those advantages and the potential applications of PQCs on near-term quantum devices have stimulated a large number of related algorithms and experiments}, including variational quantum eigensolvers
(VQE)~\cite{peruzzo2014variational}, the quantum approximate optimization algorithm (QAOA)~\cite{farhi2014quantum}, quantum generative adversarial networks~\cite{dallaire2018quantum, lloyd2018quantum,HuangPan2020,HuSun2019,HuangFan2020}, and quantum classifiers~\cite{schuld2019quantum, havlivcek2019supervised}

For QML to solve real world problems, the first step is to translate classical data, which is usually represented as a multi-dimensional array, into a quantum state. A standard way is to use a kernel function to map each element of the array into a single-qubit state, which is often referred to as qubit-encoding. The kernel function could, for example, be chosen as
\begin{align} \label{eq:kernal}
\theta_j \rightarrow \cos(\theta_j) \vert 0\rangle + \sin(\theta_j) \vert 1 \rangle,
\end{align}
which maps a real number into a single-qubit quantum state.
For an input array of size $L$, the mapping would result in an $L$-qubit quantum state, which lives in a Hilbert space of size $2^L$. For real world data with a large size, this mapping would soon become impractical for current quantum computers with less than $100$ qubits. The same problem also exists in classical deep learning, which is often built from interlacing layers of linear and nonlinear functions. A straightforward way to implement the linear function is the so-called fully connected layer, which can be represented as a dense matrix connecting each neuron of the output to all the neurons of the input. When the input size is large, this approach would become inefficient due to the large matrix size. Convolutional Neural Network (CNN)~\cite{fukushima1980neocognitron, lecun1999object} is a very popular scheme which tries to solve this problem by replacing the fully connected layer with a convolutional layer. The convolutional layer only connects each neuron of the output to a small region (window) of the input which is referred to as a feature map, thus greatly reducing the number of parameters. CNN has demonstrated itself as one of the most successful tools in the area of computer vision~\cite{NIPS2012_4824, russakovsky2015imagenet, simonyan2014very, he2016deep, goodfellow2014generative}, and more recently, it also found applications in Natural Language Processing~\cite{dauphin2016language, gehring2017convolutional, zhang2015sensitivity, kirillov2015generic, song2019abstractive, yu2018qanet}.

Inspired by CNN, we propose a hybrid quantum-classical Convolutional Neural Network ($\QCCNN$). We note that recently, a pure quantum analogy of CNN, named QCNN, was proposed to solve certain quantum many-body problems~\cite{CongLukin2019}. Similar to other QML algorithms, QCNN uses as many qubits as the size of the input, which makes it unlikely to be implemented on current quantum computers to solve real world problems. The central idea of $\QCCNN$ is to implement the feature map in the convolutional layer with a parametric quantum circuit, and correspondingly, the output of this feature map is a correlational measurement on the output quantum state of the parametric quantum circuit. In the following, we refer to this new structure as a quantum convolutional layer. As a result, the number of qubits required by this approach is only related to the window size of the feature map, which often ranges from $3\times 3$ to $9\times 9$ and is well within the capability of current quantum computers.  Moreover, since the output of our quantum convolutional layer is a classical array, it is straightforward to adapt the multi-layer structure as in CNN. Therefore, our $\QCCNN$ could utilize all the features of classical CNN, and at the same time, it is able to utilize the power of current NISQ computers. In addition, we propose a framework to automatically compute the gradients of arbitrary hybrid quantum-classical loss functions using a hybrid quantum-classical computer. To demonstrate the representative power of our $\QCCNN$, we apply it for classifying the synthetic Tetris dataset and compare its learning accuracy to classical CNN with the same architecture.

The paper is organized as follows. In Sec.~\ref{sec:arch}, we introduce our hybrid quantum-classical Convolutional Neural Network architecture. In Sec.~\ref{sec:ad}, we show a framework which can be used to automatically compute the exact gradients of any hybrid quantum-classical neural networks. In Sec.\ref{sec:results}, we demonstrate our method with the Tetris dataset and show that it can reach a higher accuracy compared to classical CNN. Finally we conclude in Sec.~\ref{sec:summary}.

\section{Hybrid Quantum-Classical Convolutional Neural Network architecture}\label{sec:arch}

\begin{figure*} [!htbp]
\includegraphics[width=2\columnwidth]{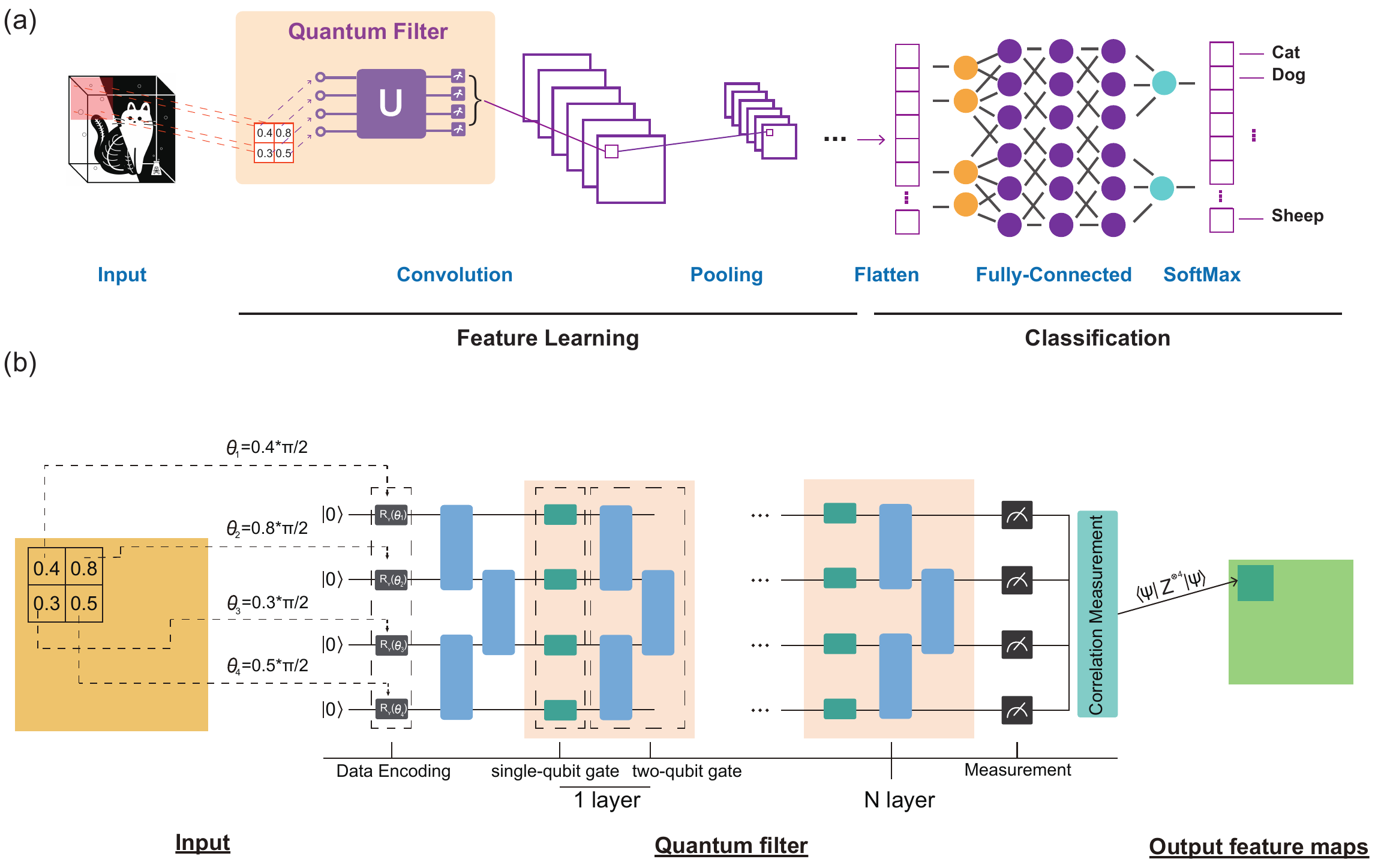}
\caption{ (a) Hybrid quantum-classical Convolutional Neural Network (QCCNN). The input demonstrated here is a two-dimensional array, which is sent to a quantum convolutional layer of $6$ filters. Each filter takes a $2\times 2$ window, translating it into a separable 4-qubit quantum state, and evolves this state with a parametric quantum circuit. For images encoded as three-dimensional arrays, the filter only works on the first two dimensions. After that a correlational measurement is made on the output quantum state and a scalar is obtained. Gathering the scalar outputs, the final output of the quantum convolutional layer is a $3$-dimensional array. Then a pooling layer is used to reduce the dimensionality of the data. This process could be repeated and finally ends with a fully connected layer. (b) Details of our design of parametric quantum circuit, which is made of interlaced single-qubit layer and two-qubit layers. The single-qubit layer consists of $\Ry$ gates, each containing one tunable parameter. The two-qubit layer consists of CNOT gates on nearest-neighbour pairs of qubits.
 \label{fig:fig1} }
\end{figure*}

To better describe our design of $\QCCNN$ (see Fig.~\ref{fig:fig1}(a)), we first briefly outline some basic features of CNN. CNN consists of interlaced convolutional layers and pooling layers, and ends with a fully connected layer. The primary purpose of the convolution layer is to extract features from the input data using a feature map (or filter), which is the most computational intensive step of CNN. After the convolutional layer, it is common to add a pooling layer to reduce the dimensionality of the data and prevent overfitting. A single filter maps small regions (windows) of the input to single neurons of the output, and is parameterized by an array $P$ which has the same shape as the window. The windows are often chosen as follows. Assuming the input is a two-dimensional array $A$ of size $v\times h$, and the predefined window size is $m\times n$, then the first window is located at the upper left conner of $A$, namely $A_{1:m,1:n}$ (here $a:b$ denotes the range from $a$ to $b$). Then the mapping is done by the linear function
\begin{align}
 A_{1:m,1:n} \rightarrow \sum_{1\leq i\leq m, 1\leq j\leq n}A_{i,j}P_{i,j}.
\end{align}
Then the next window slides to the right with a stride value $s$, which is often chosen to be $1$, until it reaches the right edge. After that it hops down to the (left) beginning of the image with the same stride value $s$ and repeats the process until the entire image is traversed. As a result, after the evolution, the output will be a two-dimensional array of size $\frac{v-m+1}{s}\times \frac{h-n+1}{s}$. Moreover, in general one could have several filters in the same layer and the input could be a three-dimensional array. For example, for a three-dimensional array of size $v\times h \times d$, and if we have $k$ filters with size $m\times n$, assuming the stride $s$, then the output array would have the shape $\frac{v-m+1}{s}\times \frac{h-n+1}{s} \times (dk)$. Generally after a convolutional layer, the output would become thinner but longer. In some situations, one would like to prevent the data to become thiner, by adding zeros around the edges of input, which is referred to as padding.

In our quantum convolutional layer, the filter is redesigned to make use of the parametric quantum circuit, which is shown in Fig.~\ref{fig:fig1}(b), and we refer to it as a quantum filter. A quantum filter takes windows of shape $m\times n$, maps them into quantum states $\vert \psi^i(A_{p:(p+m-1),q:(q+n-1)}\rangle$ of $N=mn$ qubits using Eq.(\ref{eq:kernal}), and then evolves the quantum state with the parametric quantum circuit $\pcircuit$, such that the output quantum state $\vert \psi^o\rangle$ is
\begin{align}\label{eq:circuitevo}
\vert \psi^o\rangle = \pcircuit \vert \psi^i(A_{p:(p+m-1),q:(q+n-1)}\rangle.
\end{align}
After the evolution, we take the expectation value of the observable $\sgz$, thus the feature map can be written as
\begin{align}\label{eq:qfeature}
A_{p:(p+m-1),q:(q+n-1)} \rightarrow \langle \psi^o \vert \sgz \vert \psi^o \rangle.
\end{align}
Eq.(\ref{eq:qfeature}) is nonlinear and thus in our quantum convolutional layer, we do not need an additional nonlinear function such as ReLU to explicitly bring in nonlinearity. It is also clear from Eq.(\ref{eq:qfeature}) that in our approach, the minimal number of qubits required is equal to the window size. For the next window of our quantum convolutional layer, these qubits could be reused. Thus, the quantum convolutional layer is experimentally friendly and suitable for NISQ scenarios because only a few qubits are needed and no any additional usage of qRAM is required. And the quantum correlational measurement has the potential to better capture the cross-correlation inside each window. In our architecture, the pooling layers as well as the final fully connected layer are kept in the same way as CNN since they are computationally cheap, and can further induce nonlinearities.

The parametric quantum circuit we use contains interlaced single-qubit layers and two-qubit layers. The total number of gate operations of a parametric quantum circuit, denoted as $L$, grows only polynomially with the number of qubits $N$, namely $L\sim poly(N)$ gates. In our setup, the two-qubit layer consists of CNOT gates while the single qubit layer consists of rotational Y gates ($\Ry$), which is defined as
\begin{align}
\Ry(\theta) = \left[\begin{array}{cc}
\cos(\theta) & -\sin(\theta) \\
\sin(\theta) & \cos(\theta)
\end{array}\right].
\end{align}
Here we have used $\Ry$ gate instead of other single-qubit rotational gates since \gcc{then} we could represent the quantum state as well as the parametric quantum circuit using real numbers for simplicity.
Each layer of rotational gates is counted as one depth, as a result the total number of parameters in one parametric quantum circuit will be the window size times the circuit depth.
For optimization problems, it is usually helpful to provide the gradient of the loss function. The easiest way to approximately compute the gradient in our case is to use the finite difference method, which only requires forward evaluation of the loss function.

To this end, we stress that although in Fig.~\ref{fig:fig1} we have focused on classical input data (pictures), our $\QCCNN$ architecture could also be adapted to train quantum input data. For example, for an input quantum state on a square lattice of $n\times n$ qubits, our quantum filter could directly work on each $2\times 2$ sub-lattice without the classical-to-quantum encoding stage, while the rest of the architecture remains the same. As a result, $\QCCNN$ could be used, for example, as a variational ansatz for the ground state of quantum many-body systems, or to classify quantum states as in~\cite{CongLukin2019}.

 \begin{figure*} [htbp]
 \includegraphics[width=1.7\columnwidth]{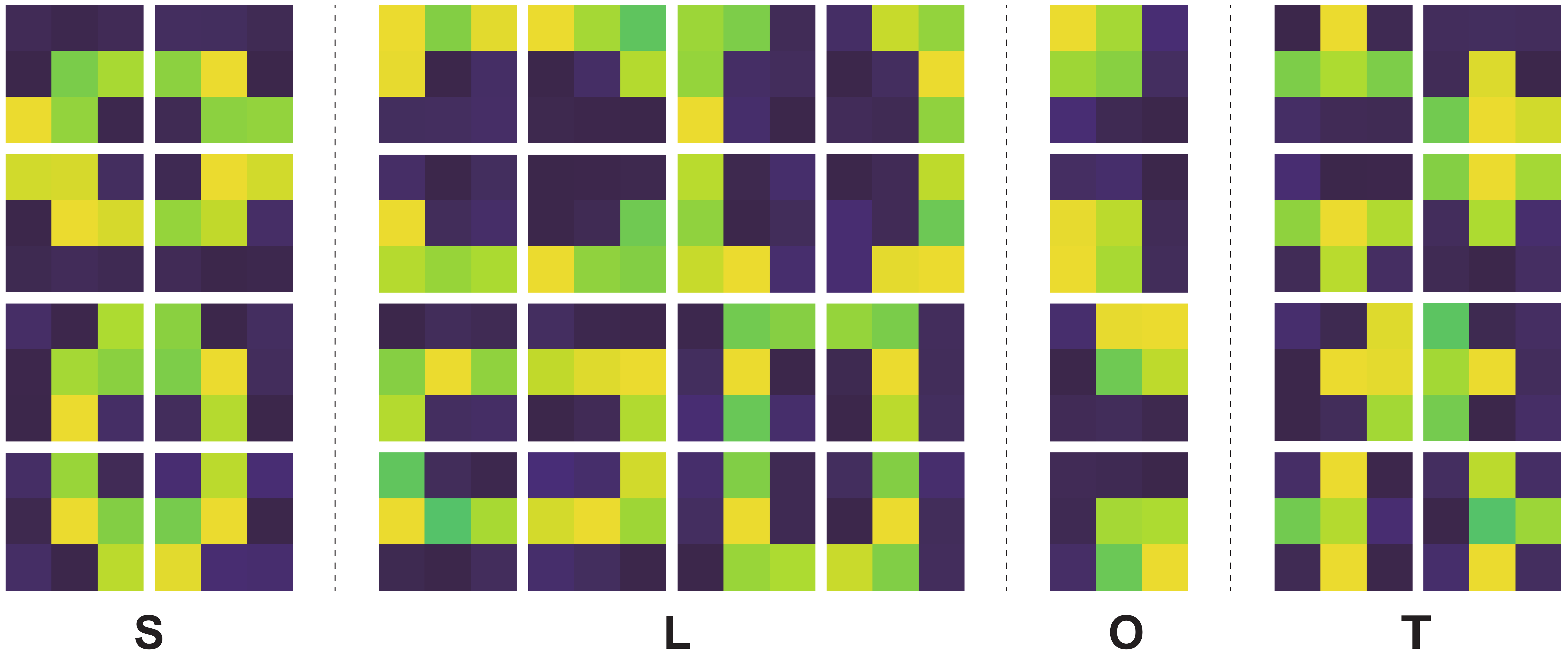}
 \caption{Some samples of the Tetris dataset. The dataset contains $1000$ gray-scale images with shape $3\times 3$. In the dataset, there are four types of Tetris bricks labeled with $S$, $L$, $O$, $T$, which have 8, 16, 4 and 8 possible configurations in the grey-scale images respectively. For each image, the foreground pixels are represented by random floating numbers ranging from $0.7$ to $1$, whereas the background are small floating numbers ranging from $0$ to $0.1$. The grey-scale images are plotted using \texttt{matplotlib.pyplot.imshow} as pseudocolor images, so that the pixels with larger numbers are represented with bright colors.
  \label{fig:fig2} }
 \end{figure*}

\section{Hybrid auto-differentiation framework}\label{sec:ad}

In the following we present a framework to compute the gradient of a hybrid quantum-classical loss function using auto-differentiation, assuming that the `quantum' sub functions involved in the loss function have the form of Eq.(\ref{eq:qfeature}).
We starting by showing how to compute the gradient of the quantum feature map, namely Eq.(\ref{eq:qfeature}).
There are two cases: i) The input $A$ is constant and the derivative against $A$ is not required and ii) The input $A$ is the output of previous steps and the derivative against $A$ is required for the following the backward propagation process. In the second case, we can add a single-qubit layer of $\Ry$ into the parametric circuit, whose parameters correspond to the values of $A$, and then the problem reduces to the first case. Therefore, it is enough to consider the gradient of the function in Eq.(\ref{eq:qfeature}), which is well-known to be~\cite{mitarai2018quantum}
\begin{align}\label{eq:qgrad}
\frac{\partial \langle \psi^o \vert \sgz \vert \psi^o \rangle}{\partial \theta_j}  =& \frac{1}{2} \left(\langle \psi^i \vert \mathcal{C}^{\dagger}(\vec{\theta}_j^+) \sgz \mathcal{C}(\vec{\theta}_j^+) \vert \psi^i \rangle \right. \nonumber \\ &\left. - \langle \psi^i \vert \mathcal{C}^{\dagger}(\vec{\theta}_j^-) \sgz \mathcal{C}(\vec{\theta}_j^-) \vert \psi^i \rangle \right),
\end{align}
where $\vec{\theta}_j^{\pm}$ means to shift the $j$-th parameter of $\vec{\theta}$, $\theta_j$ by $\pm\frac{\pi}{2}$ respectively. Now we assume for simplicity that there is a loss function $F(\vec{\theta})$ which can be written as
\begin{align}\label{eq:simpleloss}
F(\vec{\theta}) = f\left(\langle \psi^o\vert Z^{\times N}\vert \psi^o\rangle \right),
\end{align}
where $f$ is some classical function. One may attempt to compute the gradient of $F(\vec{\theta})$ by computing $F(\vec{\theta}_j^+)$ and $F(\vec{\theta}_j^-)$ and then do $\partial F(\vec{\theta}) /\partial \theta_j = (F(\vec{\theta}_j^+) - F(\vec{\theta}_j^-))/2$ similar to Eq.(\ref{eq:qgrad}), which however is incorrect in general except a few cases where $f$ is extremely simple. Other than the way to analytically derive the gradient of Eq.(\ref{eq:simpleloss}) by hand, a correct and elegant way is to integrate it into the auto-differentiation framework, for which one needs to provide the adjoint function for Eq.(\ref{eq:qfeature}) as a standard practice to extend auto-differentiation to a user-defined function. The adjoint function of Eq.(\ref{eq:qfeature}) is
\begin{align}\label{eq:adjoint}
z \rightarrow z \times \frac{\partial \langle \psi^o \vert \sgz \vert \psi^o \rangle}{\partial \theta_j},
\end{align}
which maps an input scalar $z$ to an output array with the same size as $\vec{\theta}$. Simply speaking, Eq.(\ref{eq:adjoint}) takes as input the gradient $z$ from the outer function $f$ and multiplies it with the gradient of the current function, the output of which is further passed to its inner function if there exists. Once the correct adjoint functions for each of the elementary functions of the loss function have been defined, the \gcc{computer} will be able to derive the gradient for the loss function using the auto-differentiation framework~\cite{rumelhart1988learning,GuoPoletti2020a}. The only difference here is that the adjoint function in Eq.(\ref{eq:adjoint}) depends on the output from a quantum computer. Therefore one needs to embed Eq.(\ref{eq:adjoint}) into the back-propagation process, just as one embeds Eq.(\ref{eq:qfeature}) as a sub function into the classical neural network. Then the gradient of a hybrid quantum-classical loss function would be able to be derived automatically using a hybrid quantum-classical computer. In this work the whole process is simulated using a classical computer.

Training our hybrid quantum-classical CNNs works in the same way as a regular neural network. Various gradient-based optimization techniques, such as stochastic, batch, or mini-batch gradient descent algorithms, can be used to optimize the parameters of the hybrid quantum-classical CNNs. Once the model has been trained, it can be then used to predict outputs for given inputs.

\begin{figure} [htbp]
\includegraphics[width=\columnwidth]{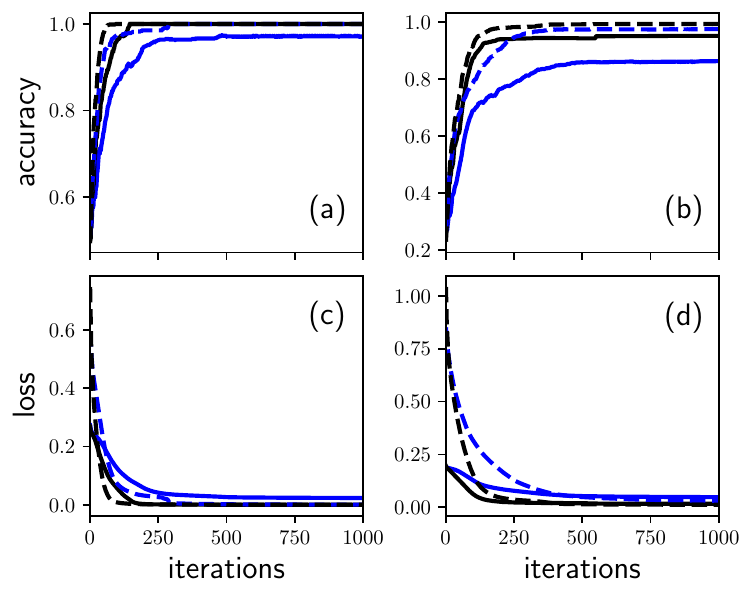}
\caption{ Accuracy and loss as a function of the number of iterations. In all the figures the blue line represents the result of one-layer CNN and the black line represents the result of two-layer CNN, the blue dashed line represents the result of one-layer QCCNN and the black dashed line represents the result of two-layer QCCNN. We have used the optimizer ADAM~\cite{kingma2014adam} and an initial learning rate of $0.01$ \gcc{and $1000$ iterations}. The results are averaged over $10$ random simulations. (a) Accuracy in case of $2$ classes. (b) Accuracy in case of $4$ classes. (c) Loss in case of $2$ classes. (d) Loss in case of $4$ classes.
 \label{fig:fig3} }
\end{figure}

\section{Numerical results and discussions}\label{sec:results}

We demonstrate the potential of our QCCNN by applying it to the Tetris dataset. We create a Tetris image dataset that consists of $800$ grey-scale images with shape $3\times 3$, in which each grey-scale image is a simulated Tetris brick (refer to a Fig.~\ref{fig:fig2} for some samples).  Concretely, the foreground pixels are represented by random floating numbers ranging from $0.7$ to $1$, whereas the background are small floating numbers ranging from $0$ to $0.1$. There are $4$ classes, \gcc{namely \textit{S}, \textit{L}, \textit{O}, and \textit{T}}, each of which represents a type of Tetris bricks. The dataset is further processed by randomly splitting into a training set and a testing set that contain 80\% and 20\% of the images, respectively. We benchmark our $\QCCNN$ against CNN with two particular structures, namely one with a single convolutional layer and another with two convolutional layers. To see the performances with a different number of classes, we create another dataset by only picking the two classes S, T out of the original training and testing data. For the single-layer structure, we use a single (quantum) convolutional layer with $5$ filters with no padding, plus a pooling layer also with no padding. For the two-layer structure, we use two (quantum) convolutional layers with $2$ and $3$ filters respectively, plus a pooling layer in between with padding $1$. The window shape for all the layers is $2\times 2$, and the stride value $s=1$. Therefore the number of qubits fed to the quantum filter is 4. \gcc{The depth of each parametric quantum circuit is set to $4$ to ensure high expressibility, as a result the total number of parameters in a quantum feature map is 16. In comparison, the number of parameters in a classical feature map is $4$, which is the same as the window size. The effect of different circuit depths in a quantum feature map will be discussed later.} 
\gcc{We use the mean square loss as our loss function $F^q$, defined as
\begin{align}\label{eq:loss}
F^{q}(\vec{a}) = \frac{1}{N_{train}} \sum_{j=1}^{N_{train}} \left(f^{q}(\vec{x}_j) - \vec{y}_j \right)^2 .
\end{align}
Here $\vec{a}$ is a list of all the parameters, $f^{q}$ means the QCCNN which is dependent on $\vec{a}$ and outputs a vector whose size is equal to the number of labels, $\vec{x}_j$ represents the $j$-th input image, $y_j$ is an integer which represents the label corresponding to $\vec{x}_j$, and $\vec{y}_j$ is the one-hot vector converted from $y_j$. $N_{train}$ is the size of the training set. The accuracy $A^q$ on the testing set is defined as
\begin{align}\label{eq:accuracy}
A^q(\vec{a}) = \frac{1}{N_{test}} \sum_{j=1}^{N_{test}} \left({\rm argmax}(f^{q}(\vec{x}_j)) == y_j\right),
\end{align}
where the function ${\rm argmax}(\vec{v})$ outputs the position of the largest element in the input vector $\vec{v}$, and $N_{test}$ is the size of the testing set. The loss function and accuracy in the classical case are defined in the same way, with QCCNN replaced by CNN in $f^q$.}
During $1000$ iterations, we compute the accuracy on the the testing data and store the values of the loss function, which is chosen as mean square loss. In Fig.~\ref{fig:fig3}(a,c), we plot the accuracy and loss values for the $2$-label case. While in Fig.~\ref{fig:fig3}(b,d) we plot the accuracy and loss values for the $4$-label case. We can see that $\QCCNN$ can reach almost $100\%$ accuracy for both the two structures we have used, and it can reach much lower loss values for both cases compared to its classical counterpart. Benefiting from the high-dimensional nature of the quantum system, the advantages of QCCNN become more transparent when the number of classes increases from $2$ to $4$. We can also see that the $4$-label case takes more iterations to converge than the $2$-label case, and that $\QCCNN$ with a two-layer structure converges faster than the single-layer structure, especially in the $4$-label case, which indicates that for complex problems, better performance could be achieved by deeper architectures.


\begin{figure} [htbp]
\includegraphics[width=\columnwidth]{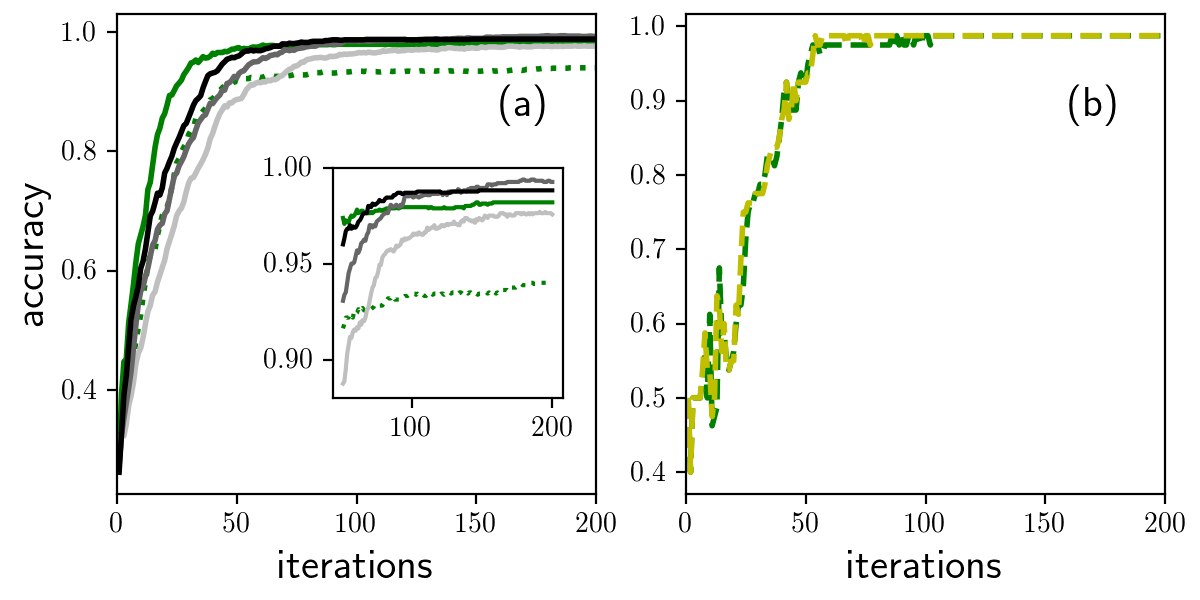}
\caption{ \gcc{(a) Accuracy as a function of the number of iterations for the two-layer QCCNN with quantum feature maps of depths $1, 2, 4$, which are shown in gray lines from lighter to darker, and for the two-layer CNN which is shown in dotted green line. The solid green line represents the accuracies computed with a modified two-layer CNN, in which the first convolutional layer uses a window size $3\times 3$ with padding $1$, followed by a pooling layer also with a window size $3\times 3$ and padding $1$. The rest of the modified two-layer CNN keeps the same as before. The inset shows the accuracies starting from the $50$-th iteration. The accuracies are averaged over $10$ random initializations and four output labels are used. (b) Accuracy as a function of the number of iterations for the noisy (green dashed line) and perfect (yellow dashed line) one-layer QCCNN with quantum feature map of depth $2$. Here $2$ output labels are used, and a single initialization with random seed $100$ is simulated due to the simulation efficiency in case of noisy quantum circuits and limited computational resources. We have assumed a single-qubit gate operation time of $100$ $ns$, a CNOT gate operation time of $300$ $ns$, and a measurement time of $1000$ $ns$. The decoherence time $T_1$ and $T_2$ for the $4$ qubits are randomly chosen from normal distributions with mean $50$ $\mu s$ and $70$ $\mu s$ respectively. We have used the ADAM optimizer with an initial learn rate $0.05$ and $200$ iterations.}
 \label{fig:fig4} }
\end{figure}

\gcc{To gain deeper insight into our QCCNN, we tune the depth of the parametric quantum circuit (thus changing the number of parameters) to see its effect on the training accuracy. Concretely, for the two-layer QCCNN, we change the circuit depth to be $1,2,4$ (the number of parameters in each quantum feature map is thus $4, 8, 16$), and the training accuracies are show in Fig.~\ref{fig:fig4}(a) with gray lines from lighter to darker. For comparison, in Fig.~\ref{fig:fig4}(a) we also show the simulation results of the corresponding two-layer CNN in green dotted line and a modified two-layer CNN in green solid line for which the window size of the first layer is enlarged to $3\times 3$ such that we could have more parameters. We can see that the QCCNN with $1$-depth quantum circuit could already reach a higher accuracy than CNN (they contain exactly the same number of parameters), and that QCCNN performs better with larger circuit depth, which is reasonable since it needs enough depth to fully entangle all the qubits. The performance of CNN could also be better when the number of parameters increase since it could explore a larger space in the first layer (space of dimension $9$ instead of $4$), which however is still slightly worse than QCCNN with circuit depths $2$ and $4$. Since current quantum devices subject to noises, we also study the effects of noisy quantum circuits on the training performance. Concretely, in Fig.~\ref{fig:fig4}(b) we show the training accuracy for a perfect (yellow dashed line) and noisy (green dashed line) one-layer QCCNN, where in the noisy quantum circuits we have considered the decoherence time $T_1$, $T_2$ as well as the measurement errors. We can see that the performance is quite resilient to such errors, and a training accuracy $1$ could still be reached in the noisy case.

Till now we have mainly focused on the training accuracy of QCCNN, and we have shown that QCCNN could potentially achieve a higher classification accuracy than CNN. We note that for the same window size $N$, the quantum feature map requires $O(Nd)$ gate operations for a parametric quantum circuit of depth $d$, while the classical feature map requires $O(N)$ Floating-point arithmetics (vector inner product). As a result from the point of computational efficiency QCCNN would not be advantageous. We could also use amplitude encoding instead of qubit encoding in principle, such that a parametric quantum circuit of $N$ qubits could process $2^N$ features in a single quantum feature map. However if naively implemented this approach would have the input problem that the encoding stage itself has a complexity of $O(2^N)$, which is already the same as the classical feature map (Our quantum feature map does not have the output problem since it outputs a single expectation value). If a better problem could be found which is free of such input problem, one could probably achieve quantum advantage in terms of computational time with our QCCNN framework.
}

\section{Conclusion}\label{sec:summary}
In summary, we present a hybrid quantum-classical Convolutional Neural Network which could be used to solve real world problems with current quantum computers. \hhl{$\QCCNN$ avoids the input and output problems of quantum machine learning algorithms when applied to real-world problems, which is one of the major challenges for current quantum machine learning algorithms.} As a quantum machine learning algorithm inspired by classical CNN, $\QCCNN$ keeps the features of CNN such as the nonlinearity, locality of the convolutional layer, as well as extensibility to deep structures. \hhl{Similar to CNN, our QCCNN architecture provides a framework for developing various hybrid quantum-classical machine learning applications on near-term quantum devices, such as discriminative model and generative model.} Moreover, the generalized feature map with a parametric quantum circuit is able to explore the correlations of neighbouring data points in a exponentially large linear space, hopefully allowing our algorithm to capture the patterns in the dataset more precisely with a quantum computer.
In addition, we also present a framework to automatically compute the gradients of hybrid quantum-classical loss functions, which could be a useful tool for developing complicated hybrid quantum-classical variational algorithms in the future. We demonstrate our approach on the Tetris dataset and show the potential of our approach to reach better learning precision for classification problems.

\textit{Note added}. During the preparation of this work, we notice a similar work that uses qRAM~\cite{KerenidisPrakash2019} and another work that uses non-parametric random quantum circuits for feature mapping~\cite{henderson2020quanvolutional}, which were carried out independently.

\begin{acknowledgments}
\textbf{Acknowledgments.}~The numerical simulation is done by the open source variational quantum circuit simulator VQC~\cite{VQC}. The noisy quantum circuits are simulated using qiskit~\cite{qiskit}. C.G. acknowledges support from National Natural Science Foundation of China under Grants No. 11805279. H.-L. H. acknowledges support from the Open Research Fund from State Key Laboratory of High Performance Computing of China (Grant No. 201901-01), National Natural Science Foundation of China under Grants No. 11905294, and China Postdoctoral Science Foundation.
\end{acknowledgments}

\bibliographystyle{apsrev4-1}
\bibliography{refs}

\begin{thebibliography}{59}%
\makeatletter
\providecommand \@ifxundefined [1]{%
 \@ifx{#1\undefined}
}%
\providecommand \@ifnum [1]{%
 \ifnum #1\expandafter \@firstoftwo
 \else \expandafter \@secondoftwo
 \fi
}%
\providecommand \@ifx [1]{%
 \ifx #1\expandafter \@firstoftwo
 \else \expandafter \@secondoftwo
 \fi
}%
\providecommand \natexlab [1]{#1}%
\providecommand \enquote  [1]{``#1''}%
\providecommand \bibnamefont  [1]{#1}%
\providecommand \bibfnamefont [1]{#1}%
\providecommand \citenamefont [1]{#1}%
\providecommand \href@noop [0]{\@secondoftwo}%
\providecommand \href [0]{\begingroup \@sanitize@url \@href}%
\providecommand \@href[1]{\@@startlink{#1}\@@href}%
\providecommand \@@href[1]{\endgroup#1\@@endlink}%
\providecommand \@sanitize@url [0]{\catcode `\\12\catcode `\$12\catcode
  `\&12\catcode `\#12\catcode `\^12\catcode `\_12\catcode `\%12\relax}%
\providecommand \@@startlink[1]{}%
\providecommand \@@endlink[0]{}%
\providecommand \url  [0]{\begingroup\@sanitize@url \@url }%
\providecommand \@url [1]{\endgroup\@href {#1}{\urlprefix }}%
\providecommand \urlprefix  [0]{URL }%
\providecommand \Eprint [0]{\href }%
\providecommand \doibase [0]{http://dx.doi.org/}%
\providecommand \selectlanguage [0]{\@gobble}%
\providecommand \bibinfo  [0]{\@secondoftwo}%
\providecommand \bibfield  [0]{\@secondoftwo}%
\providecommand \translation [1]{[#1]}%
\providecommand \BibitemOpen [0]{}%
\providecommand \bibitemStop [0]{}%
\providecommand \bibitemNoStop [0]{.\EOS\space}%
\providecommand \EOS [0]{\spacefactor3000\relax}%
\providecommand \BibitemShut  [1]{\csname bibitem#1\endcsname}%
\let\auto@bib@innerbib\@empty
\bibitem [{\citenamefont {Preskill}(2018)}]{preskill2018quantum}%
  \BibitemOpen
  \bibfield  {author} {\bibinfo {author} {\bibfnamefont {J.}~\bibnamefont
  {Preskill}},\ }\href@noop {} {\bibfield  {journal} {\bibinfo  {journal}
  {Quantum}\ }\textbf {\bibinfo {volume} {2}},\ \bibinfo {pages} {79} (\bibinfo
  {year} {2018})}\BibitemShut {NoStop}%
\bibitem [{\citenamefont {Huang}\ \emph
  {et~al.}(2020{\natexlab{a}})\citenamefont {Huang}, \citenamefont {Wu},
  \citenamefont {Fan},\ and\ \citenamefont {Zhu}}]{huang2020superconducting}%
  \BibitemOpen
  \bibfield  {author} {\bibinfo {author} {\bibfnamefont {H.-L.}\ \bibnamefont
  {Huang}}, \bibinfo {author} {\bibfnamefont {D.}~\bibnamefont {Wu}}, \bibinfo
  {author} {\bibfnamefont {D.}~\bibnamefont {Fan}}, \ and\ \bibinfo {author}
  {\bibfnamefont {X.}~\bibnamefont {Zhu}},\ }\href@noop {} {\bibfield
  {journal} {\bibinfo  {journal} {Science China Information Sciences}\ }\textbf
  {\bibinfo {volume} {63}},\ \bibinfo {pages} {1} (\bibinfo {year}
  {2020}{\natexlab{a}})}\BibitemShut {NoStop}%
\bibitem [{\citenamefont {Arute}\ \emph {et~al.}(2019)\citenamefont {Arute},
  \citenamefont {Arya}, \citenamefont {Babbush}, \citenamefont {Bacon},
  \citenamefont {Bardin}, \citenamefont {Barends}, \citenamefont {Biswas},
  \citenamefont {Boixo}, \citenamefont {Brandao}, \citenamefont {Buell} \emph
  {et~al.}}]{arute2019quantum}%
  \BibitemOpen
  \bibfield  {author} {\bibinfo {author} {\bibfnamefont {F.}~\bibnamefont
  {Arute}}, \bibinfo {author} {\bibfnamefont {K.}~\bibnamefont {Arya}},
  \bibinfo {author} {\bibfnamefont {R.}~\bibnamefont {Babbush}}, \bibinfo
  {author} {\bibfnamefont {D.}~\bibnamefont {Bacon}}, \bibinfo {author}
  {\bibfnamefont {J.~C.}\ \bibnamefont {Bardin}}, \bibinfo {author}
  {\bibfnamefont {R.}~\bibnamefont {Barends}}, \bibinfo {author} {\bibfnamefont
  {R.}~\bibnamefont {Biswas}}, \bibinfo {author} {\bibfnamefont
  {S.}~\bibnamefont {Boixo}}, \bibinfo {author} {\bibfnamefont {F.~G.}\
  \bibnamefont {Brandao}}, \bibinfo {author} {\bibfnamefont {D.~A.}\
  \bibnamefont {Buell}},  \emph {et~al.},\ }\href@noop {} {\bibfield  {journal}
  {\bibinfo  {journal} {Nature}\ }\textbf {\bibinfo {volume} {574}},\ \bibinfo
  {pages} {505} (\bibinfo {year} {2019})}\BibitemShut {NoStop}%
\bibitem [{\citenamefont {Harrow}\ and\ \citenamefont
  {Montanaro}(2017)}]{harrow2017quantum}%
  \BibitemOpen
  \bibfield  {author} {\bibinfo {author} {\bibfnamefont {A.~W.}\ \bibnamefont
  {Harrow}}\ and\ \bibinfo {author} {\bibfnamefont {A.}~\bibnamefont
  {Montanaro}},\ }\href@noop {} {\bibfield  {journal} {\bibinfo  {journal}
  {Nature}\ }\textbf {\bibinfo {volume} {549}},\ \bibinfo {pages} {203}
  (\bibinfo {year} {2017})}\BibitemShut {NoStop}%
\bibitem [{\citenamefont {Guo}\ \emph {et~al.}(2019)\citenamefont {Guo},
  \citenamefont {Liu}, \citenamefont {Xiong}, \citenamefont {Xue},
  \citenamefont {Fu}, \citenamefont {Huang}, \citenamefont {Qiang},
  \citenamefont {Xu}, \citenamefont {Liu}, \citenamefont {Zheng} \emph
  {et~al.}}]{GuoWu2019}%
  \BibitemOpen
  \bibfield  {author} {\bibinfo {author} {\bibfnamefont {C.}~\bibnamefont
  {Guo}}, \bibinfo {author} {\bibfnamefont {Y.}~\bibnamefont {Liu}}, \bibinfo
  {author} {\bibfnamefont {M.}~\bibnamefont {Xiong}}, \bibinfo {author}
  {\bibfnamefont {S.}~\bibnamefont {Xue}}, \bibinfo {author} {\bibfnamefont
  {X.}~\bibnamefont {Fu}}, \bibinfo {author} {\bibfnamefont {A.}~\bibnamefont
  {Huang}}, \bibinfo {author} {\bibfnamefont {X.}~\bibnamefont {Qiang}},
  \bibinfo {author} {\bibfnamefont {P.}~\bibnamefont {Xu}}, \bibinfo {author}
  {\bibfnamefont {J.}~\bibnamefont {Liu}}, \bibinfo {author} {\bibfnamefont
  {S.}~\bibnamefont {Zheng}},  \emph {et~al.},\ }\href@noop {} {\bibfield
  {journal} {\bibinfo  {journal} {Physical Review Letters}\ }\textbf {\bibinfo
  {volume} {123}},\ \bibinfo {pages} {190501} (\bibinfo {year}
  {2019})}\BibitemShut {NoStop}%
\bibitem [{\citenamefont {Guo}\ \emph {et~al.}(2021)\citenamefont {Guo},
  \citenamefont {Zhao},\ and\ \citenamefont {Huang}}]{guo2021verifying}%
  \BibitemOpen
  \bibfield  {author} {\bibinfo {author} {\bibfnamefont {C.}~\bibnamefont
  {Guo}}, \bibinfo {author} {\bibfnamefont {Y.}~\bibnamefont {Zhao}}, \ and\
  \bibinfo {author} {\bibfnamefont {H.-L.}\ \bibnamefont {Huang}},\ }\href@noop
  {} {\bibfield  {journal} {\bibinfo  {journal} {Physical Review Letters}\
  }\textbf {\bibinfo {volume} {126}},\ \bibinfo {pages} {070502} (\bibinfo
  {year} {2021})}\BibitemShut {NoStop}%
\bibitem [{\citenamefont {Pan}\ and\ \citenamefont
  {Zhang}(2021)}]{PanZhang2021}%
  \BibitemOpen
  \bibfield  {author} {\bibinfo {author} {\bibfnamefont {F.}~\bibnamefont
  {Pan}}\ and\ \bibinfo {author} {\bibfnamefont {P.}~\bibnamefont {Zhang}},\
  }\href@noop {} {\bibfield  {journal} {\bibinfo  {journal} {arXiv preprint
  arXiv:2103.03074}\ } (\bibinfo {year} {2021})}\BibitemShut {NoStop}%
\bibitem [{\citenamefont {Biamonte}\ \emph {et~al.}(2017)\citenamefont
  {Biamonte}, \citenamefont {Wittek}, \citenamefont {Pancotti}, \citenamefont
  {Rebentrost}, \citenamefont {Wiebe},\ and\ \citenamefont
  {Lloyd}}]{biamonte2017quantum}%
  \BibitemOpen
  \bibfield  {author} {\bibinfo {author} {\bibfnamefont {J.}~\bibnamefont
  {Biamonte}}, \bibinfo {author} {\bibfnamefont {P.}~\bibnamefont {Wittek}},
  \bibinfo {author} {\bibfnamefont {N.}~\bibnamefont {Pancotti}}, \bibinfo
  {author} {\bibfnamefont {P.}~\bibnamefont {Rebentrost}}, \bibinfo {author}
  {\bibfnamefont {N.}~\bibnamefont {Wiebe}}, \ and\ \bibinfo {author}
  {\bibfnamefont {S.}~\bibnamefont {Lloyd}},\ }\href@noop {} {\bibfield
  {journal} {\bibinfo  {journal} {Nature}\ }\textbf {\bibinfo {volume} {549}},\
  \bibinfo {pages} {195} (\bibinfo {year} {2017})}\BibitemShut {NoStop}%
\bibitem [{\citenamefont {Lloyd}\ \emph {et~al.}(2014)\citenamefont {Lloyd},
  \citenamefont {Mohseni},\ and\ \citenamefont
  {Rebentrost}}]{lloyd2014quantum}%
  \BibitemOpen
  \bibfield  {author} {\bibinfo {author} {\bibfnamefont {S.}~\bibnamefont
  {Lloyd}}, \bibinfo {author} {\bibfnamefont {M.}~\bibnamefont {Mohseni}}, \
  and\ \bibinfo {author} {\bibfnamefont {P.}~\bibnamefont {Rebentrost}},\
  }\href@noop {} {\bibfield  {journal} {\bibinfo  {journal} {Nature Physics}\
  }\textbf {\bibinfo {volume} {10}},\ \bibinfo {pages} {631} (\bibinfo {year}
  {2014})}\BibitemShut {NoStop}%
\bibitem [{\citenamefont {Lloyd}\ \emph {et~al.}(2016)\citenamefont {Lloyd},
  \citenamefont {Garnerone},\ and\ \citenamefont {Zanardi}}]{lloyd2016quantum}%
  \BibitemOpen
  \bibfield  {author} {\bibinfo {author} {\bibfnamefont {S.}~\bibnamefont
  {Lloyd}}, \bibinfo {author} {\bibfnamefont {S.}~\bibnamefont {Garnerone}}, \
  and\ \bibinfo {author} {\bibfnamefont {P.}~\bibnamefont {Zanardi}},\
  }\href@noop {} {\bibfield  {journal} {\bibinfo  {journal} {Nature
  Communications}\ }\textbf {\bibinfo {volume} {7}},\ \bibinfo {pages} {1}
  (\bibinfo {year} {2016})}\BibitemShut {NoStop}%
\bibitem [{\citenamefont {Rebentrost}\ \emph {et~al.}(2014)\citenamefont
  {Rebentrost}, \citenamefont {Mohseni},\ and\ \citenamefont
  {Lloyd}}]{rebentrost2014quantum}%
  \BibitemOpen
  \bibfield  {author} {\bibinfo {author} {\bibfnamefont {P.}~\bibnamefont
  {Rebentrost}}, \bibinfo {author} {\bibfnamefont {M.}~\bibnamefont {Mohseni}},
  \ and\ \bibinfo {author} {\bibfnamefont {S.}~\bibnamefont {Lloyd}},\
  }\href@noop {} {\bibfield  {journal} {\bibinfo  {journal} {Physical Review
  Letters}\ }\textbf {\bibinfo {volume} {113}},\ \bibinfo {pages} {130503}
  (\bibinfo {year} {2014})}\BibitemShut {NoStop}%
\bibitem [{\citenamefont {Huang}\ \emph {et~al.}(2018)\citenamefont {Huang},
  \citenamefont {Wang}, \citenamefont {Rohde}, \citenamefont {Luo},
  \citenamefont {Zhao}, \citenamefont {Liu}, \citenamefont {Li}, \citenamefont
  {Liu}, \citenamefont {Lu},\ and\ \citenamefont
  {Pan}}]{huang2018demonstration}%
  \BibitemOpen
  \bibfield  {author} {\bibinfo {author} {\bibfnamefont {H.-L.}\ \bibnamefont
  {Huang}}, \bibinfo {author} {\bibfnamefont {X.-L.}\ \bibnamefont {Wang}},
  \bibinfo {author} {\bibfnamefont {P.~P.}\ \bibnamefont {Rohde}}, \bibinfo
  {author} {\bibfnamefont {Y.-H.}\ \bibnamefont {Luo}}, \bibinfo {author}
  {\bibfnamefont {Y.-W.}\ \bibnamefont {Zhao}}, \bibinfo {author}
  {\bibfnamefont {C.}~\bibnamefont {Liu}}, \bibinfo {author} {\bibfnamefont
  {L.}~\bibnamefont {Li}}, \bibinfo {author} {\bibfnamefont {N.-L.}\
  \bibnamefont {Liu}}, \bibinfo {author} {\bibfnamefont {C.-Y.}\ \bibnamefont
  {Lu}}, \ and\ \bibinfo {author} {\bibfnamefont {J.-W.}\ \bibnamefont {Pan}},\
  }\href@noop {} {\bibfield  {journal} {\bibinfo  {journal} {Optica}\ }\textbf
  {\bibinfo {volume} {5}},\ \bibinfo {pages} {193} (\bibinfo {year}
  {2018})}\BibitemShut {NoStop}%
\bibitem [{\citenamefont {Huang}\ \emph
  {et~al.}(2017{\natexlab{a}})\citenamefont {Huang}, \citenamefont {Zhao},
  \citenamefont {Li}, \citenamefont {Li}, \citenamefont {Du}, \citenamefont
  {Fu}, \citenamefont {Zhang}, \citenamefont {Wang},\ and\ \citenamefont
  {Bao}}]{huang2017homomorphic}%
  \BibitemOpen
  \bibfield  {author} {\bibinfo {author} {\bibfnamefont {H.-L.}\ \bibnamefont
  {Huang}}, \bibinfo {author} {\bibfnamefont {Y.-W.}\ \bibnamefont {Zhao}},
  \bibinfo {author} {\bibfnamefont {T.}~\bibnamefont {Li}}, \bibinfo {author}
  {\bibfnamefont {F.-G.}\ \bibnamefont {Li}}, \bibinfo {author} {\bibfnamefont
  {Y.-T.}\ \bibnamefont {Du}}, \bibinfo {author} {\bibfnamefont {X.-Q.}\
  \bibnamefont {Fu}}, \bibinfo {author} {\bibfnamefont {S.}~\bibnamefont
  {Zhang}}, \bibinfo {author} {\bibfnamefont {X.}~\bibnamefont {Wang}}, \ and\
  \bibinfo {author} {\bibfnamefont {W.-S.}\ \bibnamefont {Bao}},\ }\href@noop
  {} {\bibfield  {journal} {\bibinfo  {journal} {Frontiers of Physics}\
  }\textbf {\bibinfo {volume} {12}},\ \bibinfo {pages} {1} (\bibinfo {year}
  {2017}{\natexlab{a}})}\BibitemShut {NoStop}%
\bibitem [{\citenamefont {Ding}\ \emph {et~al.}(2019)\citenamefont {Ding},
  \citenamefont {Bao},\ and\ \citenamefont {Huang}}]{ding2019quantum}%
  \BibitemOpen
  \bibfield  {author} {\bibinfo {author} {\bibfnamefont {C.}~\bibnamefont
  {Ding}}, \bibinfo {author} {\bibfnamefont {T.-Y.}\ \bibnamefont {Bao}}, \
  and\ \bibinfo {author} {\bibfnamefont {H.-L.}\ \bibnamefont {Huang}},\
  }\href@noop {} {\bibfield  {journal} {\bibinfo  {journal} {arXiv preprint
  arXiv:1906.08902}\ } (\bibinfo {year} {2019})}\BibitemShut {NoStop}%
\bibitem [{\citenamefont {McClean}\ \emph {et~al.}(2016)\citenamefont
  {McClean}, \citenamefont {Romero}, \citenamefont {Babbush},\ and\
  \citenamefont {Aspuru-Guzik}}]{mcclean2016theory}%
  \BibitemOpen
  \bibfield  {author} {\bibinfo {author} {\bibfnamefont {J.~R.}\ \bibnamefont
  {McClean}}, \bibinfo {author} {\bibfnamefont {J.}~\bibnamefont {Romero}},
  \bibinfo {author} {\bibfnamefont {R.}~\bibnamefont {Babbush}}, \ and\
  \bibinfo {author} {\bibfnamefont {A.}~\bibnamefont {Aspuru-Guzik}},\
  }\href@noop {} {\bibfield  {journal} {\bibinfo  {journal} {New Journal of
  Physics}\ }\textbf {\bibinfo {volume} {18}},\ \bibinfo {pages} {023023}
  (\bibinfo {year} {2016})}\BibitemShut {NoStop}%
\bibitem [{\citenamefont {Benedetti}\ \emph {et~al.}(2019)\citenamefont
  {Benedetti}, \citenamefont {Lloyd}, \citenamefont {Sack},\ and\ \citenamefont
  {Fiorentini}}]{benedetti2019parameterized}%
  \BibitemOpen
  \bibfield  {author} {\bibinfo {author} {\bibfnamefont {M.}~\bibnamefont
  {Benedetti}}, \bibinfo {author} {\bibfnamefont {E.}~\bibnamefont {Lloyd}},
  \bibinfo {author} {\bibfnamefont {S.}~\bibnamefont {Sack}}, \ and\ \bibinfo
  {author} {\bibfnamefont {M.}~\bibnamefont {Fiorentini}},\ }\href@noop {}
  {\bibfield  {journal} {\bibinfo  {journal} {Quantum Science and Technology}\
  } (\bibinfo {year} {2019})}\BibitemShut {NoStop}%
\bibitem [{\citenamefont {Liu}\ \emph {et~al.}(2020)\citenamefont {Liu},
  \citenamefont {Wang}, \citenamefont {Xue}, \citenamefont {Huang},
  \citenamefont {Fu}, \citenamefont {Qiang}, \citenamefont {Xu}, \citenamefont
  {Huang}, \citenamefont {Deng}, \citenamefont {Guo} \emph
  {et~al.}}]{LiuWu2020}%
  \BibitemOpen
  \bibfield  {author} {\bibinfo {author} {\bibfnamefont {Y.}~\bibnamefont
  {Liu}}, \bibinfo {author} {\bibfnamefont {D.}~\bibnamefont {Wang}}, \bibinfo
  {author} {\bibfnamefont {S.}~\bibnamefont {Xue}}, \bibinfo {author}
  {\bibfnamefont {A.}~\bibnamefont {Huang}}, \bibinfo {author} {\bibfnamefont
  {X.}~\bibnamefont {Fu}}, \bibinfo {author} {\bibfnamefont {X.}~\bibnamefont
  {Qiang}}, \bibinfo {author} {\bibfnamefont {P.}~\bibnamefont {Xu}}, \bibinfo
  {author} {\bibfnamefont {H.-L.}\ \bibnamefont {Huang}}, \bibinfo {author}
  {\bibfnamefont {M.}~\bibnamefont {Deng}}, \bibinfo {author} {\bibfnamefont
  {C.}~\bibnamefont {Guo}},  \emph {et~al.},\ }\href@noop {} {\bibfield
  {journal} {\bibinfo  {journal} {Physical Review A}\ }\textbf {\bibinfo
  {volume} {101}},\ \bibinfo {pages} {052316} (\bibinfo {year}
  {2020})}\BibitemShut {NoStop}%
\bibitem [{\citenamefont {Shor}(1994)}]{Shor1994}%
  \BibitemOpen
  \bibfield  {author} {\bibinfo {author} {\bibfnamefont {P.~W.}\ \bibnamefont
  {Shor}},\ }in\ \href@noop {} {\emph {\bibinfo {booktitle} {Proceedings 35th
  annual symposium on foundations of computer science}}}\ (\bibinfo
  {organization} {Ieee},\ \bibinfo {year} {1994})\ pp.\ \bibinfo {pages}
  {124--134}\BibitemShut {NoStop}%
\bibitem [{\citenamefont {Grover}(1996)}]{grover1996fast}%
  \BibitemOpen
  \bibfield  {author} {\bibinfo {author} {\bibfnamefont {L.~K.}\ \bibnamefont
  {Grover}},\ }in\ \href@noop {} {\emph {\bibinfo {booktitle} {Proceedings of
  the twenty-eighth annual ACM symposium on Theory of computing}}}\ (\bibinfo
  {year} {1996})\ pp.\ \bibinfo {pages} {212--219}\BibitemShut {NoStop}%
\bibitem [{\citenamefont {Long}(2001)}]{long2001grover}%
  \BibitemOpen
  \bibfield  {author} {\bibinfo {author} {\bibfnamefont {G.-L.}\ \bibnamefont
  {Long}},\ }\href@noop {} {\bibfield  {journal} {\bibinfo  {journal} {Physical
  Review A}\ }\textbf {\bibinfo {volume} {64}},\ \bibinfo {pages} {022307}
  (\bibinfo {year} {2001})}\BibitemShut {NoStop}%
\bibitem [{\citenamefont {Lu}\ \emph {et~al.}(2007)\citenamefont {Lu},
  \citenamefont {Browne}, \citenamefont {Yang},\ and\ \citenamefont
  {Pan}}]{lu2007demonstration}%
  \BibitemOpen
  \bibfield  {author} {\bibinfo {author} {\bibfnamefont {C.-Y.}\ \bibnamefont
  {Lu}}, \bibinfo {author} {\bibfnamefont {D.~E.}\ \bibnamefont {Browne}},
  \bibinfo {author} {\bibfnamefont {T.}~\bibnamefont {Yang}}, \ and\ \bibinfo
  {author} {\bibfnamefont {J.-W.}\ \bibnamefont {Pan}},\ }\href@noop {}
  {\bibfield  {journal} {\bibinfo  {journal} {Physical Review Letters}\
  }\textbf {\bibinfo {volume} {99}},\ \bibinfo {pages} {250504} (\bibinfo
  {year} {2007})}\BibitemShut {NoStop}%
\bibitem [{\citenamefont {Huang}\ \emph
  {et~al.}(2017{\natexlab{b}})\citenamefont {Huang}, \citenamefont {Zhao},
  \citenamefont {Ma}, \citenamefont {Liu}, \citenamefont {Su}, \citenamefont
  {Wang}, \citenamefont {Li}, \citenamefont {Liu}, \citenamefont {Sanders},
  \citenamefont {Lu} \emph {et~al.}}]{huang2017experimental}%
  \BibitemOpen
  \bibfield  {author} {\bibinfo {author} {\bibfnamefont {H.-L.}\ \bibnamefont
  {Huang}}, \bibinfo {author} {\bibfnamefont {Q.}~\bibnamefont {Zhao}},
  \bibinfo {author} {\bibfnamefont {X.}~\bibnamefont {Ma}}, \bibinfo {author}
  {\bibfnamefont {C.}~\bibnamefont {Liu}}, \bibinfo {author} {\bibfnamefont
  {Z.-E.}\ \bibnamefont {Su}}, \bibinfo {author} {\bibfnamefont {X.-L.}\
  \bibnamefont {Wang}}, \bibinfo {author} {\bibfnamefont {L.}~\bibnamefont
  {Li}}, \bibinfo {author} {\bibfnamefont {N.-L.}\ \bibnamefont {Liu}},
  \bibinfo {author} {\bibfnamefont {B.~C.}\ \bibnamefont {Sanders}}, \bibinfo
  {author} {\bibfnamefont {C.-Y.}\ \bibnamefont {Lu}},  \emph {et~al.},\
  }\href@noop {} {\bibfield  {journal} {\bibinfo  {journal} {Physical Review
  Letters}\ }\textbf {\bibinfo {volume} {119}},\ \bibinfo {pages} {050503}
  (\bibinfo {year} {2017}{\natexlab{b}})}\BibitemShut {NoStop}%
\bibitem [{\citenamefont {Gentini}\ \emph {et~al.}(2020)\citenamefont
  {Gentini}, \citenamefont {Cuccoli}, \citenamefont {Pirandola}, \citenamefont
  {Verrucchi},\ and\ \citenamefont {Banchi}}]{gentini2020noise}%
  \BibitemOpen
  \bibfield  {author} {\bibinfo {author} {\bibfnamefont {L.}~\bibnamefont
  {Gentini}}, \bibinfo {author} {\bibfnamefont {A.}~\bibnamefont {Cuccoli}},
  \bibinfo {author} {\bibfnamefont {S.}~\bibnamefont {Pirandola}}, \bibinfo
  {author} {\bibfnamefont {P.}~\bibnamefont {Verrucchi}}, \ and\ \bibinfo
  {author} {\bibfnamefont {L.}~\bibnamefont {Banchi}},\ }\href@noop {}
  {\bibfield  {journal} {\bibinfo  {journal} {Physical Review A}\ }\textbf
  {\bibinfo {volume} {102}},\ \bibinfo {pages} {052414} (\bibinfo {year}
  {2020})}\BibitemShut {NoStop}%
\bibitem [{\citenamefont {Killoran}\ \emph {et~al.}(2019)\citenamefont
  {Killoran}, \citenamefont {Bromley}, \citenamefont {Arrazola}, \citenamefont
  {Schuld}, \citenamefont {Quesada},\ and\ \citenamefont
  {Lloyd}}]{killoran2019continuous}%
  \BibitemOpen
  \bibfield  {author} {\bibinfo {author} {\bibfnamefont {N.}~\bibnamefont
  {Killoran}}, \bibinfo {author} {\bibfnamefont {T.~R.}\ \bibnamefont
  {Bromley}}, \bibinfo {author} {\bibfnamefont {J.~M.}\ \bibnamefont
  {Arrazola}}, \bibinfo {author} {\bibfnamefont {M.}~\bibnamefont {Schuld}},
  \bibinfo {author} {\bibfnamefont {N.}~\bibnamefont {Quesada}}, \ and\
  \bibinfo {author} {\bibfnamefont {S.}~\bibnamefont {Lloyd}},\ }\href@noop {}
  {\bibfield  {journal} {\bibinfo  {journal} {Physical Review Research}\
  }\textbf {\bibinfo {volume} {1}},\ \bibinfo {pages} {033063} (\bibinfo {year}
  {2019})}\BibitemShut {NoStop}%
\bibitem [{\citenamefont {Du}\ \emph {et~al.}(2020)\citenamefont {Du},
  \citenamefont {Hsieh}, \citenamefont {Liu},\ and\ \citenamefont
  {Tao}}]{du2020expressive}%
  \BibitemOpen
  \bibfield  {author} {\bibinfo {author} {\bibfnamefont {Y.}~\bibnamefont
  {Du}}, \bibinfo {author} {\bibfnamefont {M.-H.}\ \bibnamefont {Hsieh}},
  \bibinfo {author} {\bibfnamefont {T.}~\bibnamefont {Liu}}, \ and\ \bibinfo
  {author} {\bibfnamefont {D.}~\bibnamefont {Tao}},\ }\href@noop {} {\bibfield
  {journal} {\bibinfo  {journal} {Physical Review Research}\ }\textbf {\bibinfo
  {volume} {2}},\ \bibinfo {pages} {033125} (\bibinfo {year}
  {2020})}\BibitemShut {NoStop}%
\bibitem [{\citenamefont {Coyle}\ \emph {et~al.}(2020)\citenamefont {Coyle},
  \citenamefont {Mills}, \citenamefont {Danos},\ and\ \citenamefont
  {Kashefi}}]{coyle2020born}%
  \BibitemOpen
  \bibfield  {author} {\bibinfo {author} {\bibfnamefont {B.}~\bibnamefont
  {Coyle}}, \bibinfo {author} {\bibfnamefont {D.}~\bibnamefont {Mills}},
  \bibinfo {author} {\bibfnamefont {V.}~\bibnamefont {Danos}}, \ and\ \bibinfo
  {author} {\bibfnamefont {E.}~\bibnamefont {Kashefi}},\ }\href@noop {}
  {\bibfield  {journal} {\bibinfo  {journal} {npj Quantum Information}\
  }\textbf {\bibinfo {volume} {6}},\ \bibinfo {pages} {1} (\bibinfo {year}
  {2020})}\BibitemShut {NoStop}%
\bibitem [{\citenamefont {Tangpanitanon}\ \emph {et~al.}(2020)\citenamefont
  {Tangpanitanon}, \citenamefont {Thanasilp}, \citenamefont {Dangniam},
  \citenamefont {Lemonde},\ and\ \citenamefont
  {Angelakis}}]{tangpanitanon2020expressibility}%
  \BibitemOpen
  \bibfield  {author} {\bibinfo {author} {\bibfnamefont {J.}~\bibnamefont
  {Tangpanitanon}}, \bibinfo {author} {\bibfnamefont {S.}~\bibnamefont
  {Thanasilp}}, \bibinfo {author} {\bibfnamefont {N.}~\bibnamefont {Dangniam}},
  \bibinfo {author} {\bibfnamefont {M.-A.}\ \bibnamefont {Lemonde}}, \ and\
  \bibinfo {author} {\bibfnamefont {D.~G.}\ \bibnamefont {Angelakis}},\
  }\href@noop {} {\bibfield  {journal} {\bibinfo  {journal} {Physical Review
  Research}\ }\textbf {\bibinfo {volume} {2}},\ \bibinfo {pages} {043364}
  (\bibinfo {year} {2020})}\BibitemShut {NoStop}%
\bibitem [{\citenamefont {Schuld}\ and\ \citenamefont
  {Killoran}(2019)}]{schuld2019quantum}%
  \BibitemOpen
  \bibfield  {author} {\bibinfo {author} {\bibfnamefont {M.}~\bibnamefont
  {Schuld}}\ and\ \bibinfo {author} {\bibfnamefont {N.}~\bibnamefont
  {Killoran}},\ }\href@noop {} {\bibfield  {journal} {\bibinfo  {journal}
  {Physical Review Letters}\ }\textbf {\bibinfo {volume} {122}},\ \bibinfo
  {pages} {040504} (\bibinfo {year} {2019})}\BibitemShut {NoStop}%
\bibitem [{\citenamefont {Sarma}\ \emph {et~al.}(2019)\citenamefont {Sarma},
  \citenamefont {Deng},\ and\ \citenamefont {Duan}}]{SarmaDuan2019}%
  \BibitemOpen
  \bibfield  {author} {\bibinfo {author} {\bibfnamefont {S.~D.}\ \bibnamefont
  {Sarma}}, \bibinfo {author} {\bibfnamefont {D.-L.}\ \bibnamefont {Deng}}, \
  and\ \bibinfo {author} {\bibfnamefont {L.-M.}\ \bibnamefont {Duan}},\
  }\href@noop {} {\bibfield  {journal} {\bibinfo  {journal} {arXiv preprint
  arXiv:1903.03516}\ } (\bibinfo {year} {2019})}\BibitemShut {NoStop}%
\bibitem [{\citenamefont {Lloyd}\ and\ \citenamefont
  {Weedbrook}(2018)}]{lloyd2018quantum}%
  \BibitemOpen
  \bibfield  {author} {\bibinfo {author} {\bibfnamefont {S.}~\bibnamefont
  {Lloyd}}\ and\ \bibinfo {author} {\bibfnamefont {C.}~\bibnamefont
  {Weedbrook}},\ }\href@noop {} {\bibfield  {journal} {\bibinfo  {journal}
  {Physical Review Letters}\ }\textbf {\bibinfo {volume} {121}},\ \bibinfo
  {pages} {040502} (\bibinfo {year} {2018})}\BibitemShut {NoStop}%
\bibitem [{\citenamefont {Peruzzo}\ \emph {et~al.}(2014)\citenamefont
  {Peruzzo}, \citenamefont {McClean}, \citenamefont {Shadbolt}, \citenamefont
  {Yung}, \citenamefont {Zhou}, \citenamefont {Love}, \citenamefont
  {Aspuru-Guzik},\ and\ \citenamefont {Obrien}}]{peruzzo2014variational}%
  \BibitemOpen
  \bibfield  {author} {\bibinfo {author} {\bibfnamefont {A.}~\bibnamefont
  {Peruzzo}}, \bibinfo {author} {\bibfnamefont {J.}~\bibnamefont {McClean}},
  \bibinfo {author} {\bibfnamefont {P.}~\bibnamefont {Shadbolt}}, \bibinfo
  {author} {\bibfnamefont {M.-H.}\ \bibnamefont {Yung}}, \bibinfo {author}
  {\bibfnamefont {X.-Q.}\ \bibnamefont {Zhou}}, \bibinfo {author}
  {\bibfnamefont {P.~J.}\ \bibnamefont {Love}}, \bibinfo {author}
  {\bibfnamefont {A.}~\bibnamefont {Aspuru-Guzik}}, \ and\ \bibinfo {author}
  {\bibfnamefont {J.~L.}\ \bibnamefont {Obrien}},\ }\href@noop {} {\bibfield
  {journal} {\bibinfo  {journal} {Nature Communications}\ }\textbf {\bibinfo
  {volume} {5}},\ \bibinfo {pages} {4213} (\bibinfo {year} {2014})}\BibitemShut
  {NoStop}%
\bibitem [{\citenamefont {Farhi}\ \emph {et~al.}(2014)\citenamefont {Farhi},
  \citenamefont {Goldstone},\ and\ \citenamefont {Gutmann}}]{farhi2014quantum}%
  \BibitemOpen
  \bibfield  {author} {\bibinfo {author} {\bibfnamefont {E.}~\bibnamefont
  {Farhi}}, \bibinfo {author} {\bibfnamefont {J.}~\bibnamefont {Goldstone}}, \
  and\ \bibinfo {author} {\bibfnamefont {S.}~\bibnamefont {Gutmann}},\
  }\href@noop {} {\bibfield  {journal} {\bibinfo  {journal} {arXiv:1411.4028}\
  } (\bibinfo {year} {2014})}\BibitemShut {NoStop}%
\bibitem [{\citenamefont {Dallaire-Demers}\ and\ \citenamefont
  {Killoran}(2018)}]{dallaire2018quantum}%
  \BibitemOpen
  \bibfield  {author} {\bibinfo {author} {\bibfnamefont {P.-L.}\ \bibnamefont
  {Dallaire-Demers}}\ and\ \bibinfo {author} {\bibfnamefont {N.}~\bibnamefont
  {Killoran}},\ }\href@noop {} {\bibfield  {journal} {\bibinfo  {journal}
  {Physical Review A}\ }\textbf {\bibinfo {volume} {98}},\ \bibinfo {pages}
  {012324} (\bibinfo {year} {2018})}\BibitemShut {NoStop}%
\bibitem [{\citenamefont {Huang}\ \emph
  {et~al.}(2020{\natexlab{b}})\citenamefont {Huang}, \citenamefont {Du},
  \citenamefont {Gong}, \citenamefont {Zhao}, \citenamefont {Wu}, \citenamefont
  {Wang}, \citenamefont {Li}, \citenamefont {Liang}, \citenamefont {Lin},
  \citenamefont {Xu} \emph {et~al.}}]{HuangPan2020}%
  \BibitemOpen
  \bibfield  {author} {\bibinfo {author} {\bibfnamefont {H.-L.}\ \bibnamefont
  {Huang}}, \bibinfo {author} {\bibfnamefont {Y.}~\bibnamefont {Du}}, \bibinfo
  {author} {\bibfnamefont {M.}~\bibnamefont {Gong}}, \bibinfo {author}
  {\bibfnamefont {Y.}~\bibnamefont {Zhao}}, \bibinfo {author} {\bibfnamefont
  {Y.}~\bibnamefont {Wu}}, \bibinfo {author} {\bibfnamefont {C.}~\bibnamefont
  {Wang}}, \bibinfo {author} {\bibfnamefont {S.}~\bibnamefont {Li}}, \bibinfo
  {author} {\bibfnamefont {F.}~\bibnamefont {Liang}}, \bibinfo {author}
  {\bibfnamefont {J.}~\bibnamefont {Lin}}, \bibinfo {author} {\bibfnamefont
  {Y.}~\bibnamefont {Xu}},  \emph {et~al.},\ }\href@noop {} {\bibfield
  {journal} {\bibinfo  {journal} {arXiv:2010.06201}\ } (\bibinfo {year}
  {2020}{\natexlab{b}})}\BibitemShut {NoStop}%
\bibitem [{\citenamefont {Hu}\ \emph {et~al.}(2019)\citenamefont {Hu},
  \citenamefont {Wu}, \citenamefont {Cai}, \citenamefont {Ma}, \citenamefont
  {Mu}, \citenamefont {Xu}, \citenamefont {Wang}, \citenamefont {Song},
  \citenamefont {Deng}, \citenamefont {Zou},\ and\ \citenamefont
  {Sun}}]{HuSun2019}%
  \BibitemOpen
  \bibfield  {author} {\bibinfo {author} {\bibfnamefont {L.}~\bibnamefont
  {Hu}}, \bibinfo {author} {\bibfnamefont {S.-H.}\ \bibnamefont {Wu}}, \bibinfo
  {author} {\bibfnamefont {W.}~\bibnamefont {Cai}}, \bibinfo {author}
  {\bibfnamefont {Y.}~\bibnamefont {Ma}}, \bibinfo {author} {\bibfnamefont
  {X.}~\bibnamefont {Mu}}, \bibinfo {author} {\bibfnamefont {Y.}~\bibnamefont
  {Xu}}, \bibinfo {author} {\bibfnamefont {H.}~\bibnamefont {Wang}}, \bibinfo
  {author} {\bibfnamefont {Y.}~\bibnamefont {Song}}, \bibinfo {author}
  {\bibfnamefont {D.-L.}\ \bibnamefont {Deng}}, \bibinfo {author}
  {\bibfnamefont {C.-L.}\ \bibnamefont {Zou}}, \ and\ \bibinfo {author}
  {\bibfnamefont {L.}~\bibnamefont {Sun}},\ }\href@noop {} {\bibfield
  {journal} {\bibinfo  {journal} {Science advances}\ }\textbf {\bibinfo
  {volume} {5}},\ \bibinfo {pages} {eaav2761} (\bibinfo {year}
  {2019})}\BibitemShut {NoStop}%
\bibitem [{\citenamefont {Huang}\ \emph
  {et~al.}(2020{\natexlab{c}})\citenamefont {Huang}, \citenamefont {Wang},
  \citenamefont {Song}, \citenamefont {Xu}, \citenamefont {Li}, \citenamefont
  {Wang}, \citenamefont {Guo}, \citenamefont {Song}, \citenamefont {Liu},
  \citenamefont {Zheng}, \citenamefont {Deng}, \citenamefont {Wang},
  \citenamefont {Jian-Guo},\ and\ \citenamefont {Fan}}]{HuangFan2020}%
  \BibitemOpen
  \bibfield  {author} {\bibinfo {author} {\bibfnamefont {K.}~\bibnamefont
  {Huang}}, \bibinfo {author} {\bibfnamefont {Z.-A.}\ \bibnamefont {Wang}},
  \bibinfo {author} {\bibfnamefont {C.}~\bibnamefont {Song}}, \bibinfo {author}
  {\bibfnamefont {K.}~\bibnamefont {Xu}}, \bibinfo {author} {\bibfnamefont
  {H.}~\bibnamefont {Li}}, \bibinfo {author} {\bibfnamefont {Z.}~\bibnamefont
  {Wang}}, \bibinfo {author} {\bibfnamefont {Q.}~\bibnamefont {Guo}}, \bibinfo
  {author} {\bibfnamefont {Z.}~\bibnamefont {Song}}, \bibinfo {author}
  {\bibfnamefont {Z.-B.}\ \bibnamefont {Liu}}, \bibinfo {author} {\bibfnamefont
  {D.}~\bibnamefont {Zheng}}, \bibinfo {author} {\bibfnamefont {D.-L.}\
  \bibnamefont {Deng}}, \bibinfo {author} {\bibfnamefont {H.}~\bibnamefont
  {Wang}}, \bibinfo {author} {\bibfnamefont {T.}~\bibnamefont {Jian-Guo}}, \
  and\ \bibinfo {author} {\bibfnamefont {H.}~\bibnamefont {Fan}},\ }\href@noop
  {} {\bibfield  {journal} {\bibinfo  {journal} {arXiv preprint
  arXiv:2009.12827}\ } (\bibinfo {year} {2020}{\natexlab{c}})}\BibitemShut
  {NoStop}%
\bibitem [{\citenamefont {Havl{\'\i}{\v{c}}ek}\ \emph
  {et~al.}(2019)\citenamefont {Havl{\'\i}{\v{c}}ek}, \citenamefont
  {C{\'o}rcoles}, \citenamefont {Temme}, \citenamefont {Harrow}, \citenamefont
  {Kandala}, \citenamefont {Chow},\ and\ \citenamefont
  {Gambetta}}]{havlivcek2019supervised}%
  \BibitemOpen
  \bibfield  {author} {\bibinfo {author} {\bibfnamefont {V.}~\bibnamefont
  {Havl{\'\i}{\v{c}}ek}}, \bibinfo {author} {\bibfnamefont {A.~D.}\
  \bibnamefont {C{\'o}rcoles}}, \bibinfo {author} {\bibfnamefont
  {K.}~\bibnamefont {Temme}}, \bibinfo {author} {\bibfnamefont {A.~W.}\
  \bibnamefont {Harrow}}, \bibinfo {author} {\bibfnamefont {A.}~\bibnamefont
  {Kandala}}, \bibinfo {author} {\bibfnamefont {J.~M.}\ \bibnamefont {Chow}}, \
  and\ \bibinfo {author} {\bibfnamefont {J.~M.}\ \bibnamefont {Gambetta}},\
  }\href@noop {} {\bibfield  {journal} {\bibinfo  {journal} {Nature}\ }\textbf
  {\bibinfo {volume} {567}},\ \bibinfo {pages} {209} (\bibinfo {year}
  {2019})}\BibitemShut {NoStop}%
\bibitem [{\citenamefont {Fukushima}(1980)}]{fukushima1980neocognitron}%
  \BibitemOpen
  \bibfield  {author} {\bibinfo {author} {\bibfnamefont {K.}~\bibnamefont
  {Fukushima}},\ }\href@noop {} {\bibfield  {journal} {\bibinfo  {journal}
  {Biological cybernetics}\ }\textbf {\bibinfo {volume} {36}},\ \bibinfo
  {pages} {193} (\bibinfo {year} {1980})}\BibitemShut {NoStop}%
\bibitem [{\citenamefont {LeCun}\ \emph {et~al.}(1999)\citenamefont {LeCun},
  \citenamefont {Haffner}, \citenamefont {Bottou},\ and\ \citenamefont
  {Bengio}}]{lecun1999object}%
  \BibitemOpen
  \bibfield  {author} {\bibinfo {author} {\bibfnamefont {Y.}~\bibnamefont
  {LeCun}}, \bibinfo {author} {\bibfnamefont {P.}~\bibnamefont {Haffner}},
  \bibinfo {author} {\bibfnamefont {L.}~\bibnamefont {Bottou}}, \ and\ \bibinfo
  {author} {\bibfnamefont {Y.}~\bibnamefont {Bengio}},\ }in\ \href@noop {}
  {\emph {\bibinfo {booktitle} {Shape, contour and grouping in computer
  vision}}}\ (\bibinfo  {publisher} {Springer},\ \bibinfo {year} {1999})\ pp.\
  \bibinfo {pages} {319--345}\BibitemShut {NoStop}%
\bibitem [{\citenamefont {Krizhevsky}\ \emph {et~al.}(2012)\citenamefont
  {Krizhevsky}, \citenamefont {Sutskever},\ and\ \citenamefont
  {Hinton}}]{NIPS2012_4824}%
  \BibitemOpen
  \bibfield  {author} {\bibinfo {author} {\bibfnamefont {A.}~\bibnamefont
  {Krizhevsky}}, \bibinfo {author} {\bibfnamefont {I.}~\bibnamefont
  {Sutskever}}, \ and\ \bibinfo {author} {\bibfnamefont {G.~E.}\ \bibnamefont
  {Hinton}},\ }in\ \href@noop {} {\emph {\bibinfo {booktitle} {Advances in
  Neural Information Processing Systems 25}}},\ \bibinfo {editor} {edited by\
  \bibinfo {editor} {\bibfnamefont {F.}~\bibnamefont {Pereira}}, \bibinfo
  {editor} {\bibfnamefont {C.~J.~C.}\ \bibnamefont {Burges}}, \bibinfo {editor}
  {\bibfnamefont {L.}~\bibnamefont {Bottou}}, \ and\ \bibinfo {editor}
  {\bibfnamefont {K.~Q.}\ \bibnamefont {Weinberger}}}\ (\bibinfo  {publisher}
  {Curran Associates, Inc.},\ \bibinfo {year} {2012})\ pp.\ \bibinfo {pages}
  {1097--1105}\BibitemShut {NoStop}%
\bibitem [{\citenamefont {Russakovsky}\ \emph {et~al.}(2015)\citenamefont
  {Russakovsky}, \citenamefont {Deng}, \citenamefont {Su}, \citenamefont
  {Krause}, \citenamefont {Satheesh}, \citenamefont {Ma}, \citenamefont
  {Huang}, \citenamefont {Karpathy}, \citenamefont {Khosla}, \citenamefont
  {Bernstein} \emph {et~al.}}]{russakovsky2015imagenet}%
  \BibitemOpen
  \bibfield  {author} {\bibinfo {author} {\bibfnamefont {O.}~\bibnamefont
  {Russakovsky}}, \bibinfo {author} {\bibfnamefont {J.}~\bibnamefont {Deng}},
  \bibinfo {author} {\bibfnamefont {H.}~\bibnamefont {Su}}, \bibinfo {author}
  {\bibfnamefont {J.}~\bibnamefont {Krause}}, \bibinfo {author} {\bibfnamefont
  {S.}~\bibnamefont {Satheesh}}, \bibinfo {author} {\bibfnamefont
  {S.}~\bibnamefont {Ma}}, \bibinfo {author} {\bibfnamefont {Z.}~\bibnamefont
  {Huang}}, \bibinfo {author} {\bibfnamefont {A.}~\bibnamefont {Karpathy}},
  \bibinfo {author} {\bibfnamefont {A.}~\bibnamefont {Khosla}}, \bibinfo
  {author} {\bibfnamefont {M.}~\bibnamefont {Bernstein}},  \emph {et~al.},\
  }\href@noop {} {\bibfield  {journal} {\bibinfo  {journal} {International
  journal of computer vision}\ }\textbf {\bibinfo {volume} {115}},\ \bibinfo
  {pages} {211} (\bibinfo {year} {2015})}\BibitemShut {NoStop}%
\bibitem [{\citenamefont {Simonyan}\ and\ \citenamefont
  {Zisserman}(2014)}]{simonyan2014very}%
  \BibitemOpen
  \bibfield  {author} {\bibinfo {author} {\bibfnamefont {K.}~\bibnamefont
  {Simonyan}}\ and\ \bibinfo {author} {\bibfnamefont {A.}~\bibnamefont
  {Zisserman}},\ }\href@noop {} {\bibfield  {journal} {\bibinfo  {journal}
  {arXiv:1409.1556}\ } (\bibinfo {year} {2014})}\BibitemShut {NoStop}%
\bibitem [{\citenamefont {He}\ \emph {et~al.}(2016)\citenamefont {He},
  \citenamefont {Zhang}, \citenamefont {Ren},\ and\ \citenamefont
  {Sun}}]{he2016deep}%
  \BibitemOpen
  \bibfield  {author} {\bibinfo {author} {\bibfnamefont {K.}~\bibnamefont
  {He}}, \bibinfo {author} {\bibfnamefont {X.}~\bibnamefont {Zhang}}, \bibinfo
  {author} {\bibfnamefont {S.}~\bibnamefont {Ren}}, \ and\ \bibinfo {author}
  {\bibfnamefont {J.}~\bibnamefont {Sun}},\ }in\ \href@noop {} {\emph {\bibinfo
  {booktitle} {Proceedings of the IEEE conference on computer vision and
  pattern recognition}}}\ (\bibinfo {year} {2016})\ pp.\ \bibinfo {pages}
  {770--778}\BibitemShut {NoStop}%
\bibitem [{\citenamefont {Goodfellow}\ \emph {et~al.}(2014)\citenamefont
  {Goodfellow}, \citenamefont {Pouget-Abadie}, \citenamefont {Mirza},
  \citenamefont {Xu}, \citenamefont {Warde-Farley}, \citenamefont {Ozair},
  \citenamefont {Courville},\ and\ \citenamefont
  {Bengio}}]{goodfellow2014generative}%
  \BibitemOpen
  \bibfield  {author} {\bibinfo {author} {\bibfnamefont {I.~J.}\ \bibnamefont
  {Goodfellow}}, \bibinfo {author} {\bibfnamefont {J.}~\bibnamefont
  {Pouget-Abadie}}, \bibinfo {author} {\bibfnamefont {M.}~\bibnamefont
  {Mirza}}, \bibinfo {author} {\bibfnamefont {B.}~\bibnamefont {Xu}}, \bibinfo
  {author} {\bibfnamefont {D.}~\bibnamefont {Warde-Farley}}, \bibinfo {author}
  {\bibfnamefont {S.}~\bibnamefont {Ozair}}, \bibinfo {author} {\bibfnamefont
  {A.}~\bibnamefont {Courville}}, \ and\ \bibinfo {author} {\bibfnamefont
  {Y.}~\bibnamefont {Bengio}},\ }\href@noop {} {\bibfield  {journal} {\bibinfo
  {journal} {arXiv:1406.2661}\ } (\bibinfo {year} {2014})}\BibitemShut
  {NoStop}%
\bibitem [{\citenamefont {Dauphin}\ \emph {et~al.}(2016)\citenamefont
  {Dauphin}, \citenamefont {Fan}, \citenamefont {Auli},\ and\ \citenamefont
  {Grangier}}]{dauphin2016language}%
  \BibitemOpen
  \bibfield  {author} {\bibinfo {author} {\bibfnamefont {Y.~N.}\ \bibnamefont
  {Dauphin}}, \bibinfo {author} {\bibfnamefont {A.}~\bibnamefont {Fan}},
  \bibinfo {author} {\bibfnamefont {M.}~\bibnamefont {Auli}}, \ and\ \bibinfo
  {author} {\bibfnamefont {D.}~\bibnamefont {Grangier}},\ }\href@noop {}
  {\bibfield  {journal} {\bibinfo  {journal} {arXiv:1612.08083}\ } (\bibinfo
  {year} {2016})}\BibitemShut {NoStop}%
\bibitem [{\citenamefont {Gehring}\ \emph {et~al.}(2017)\citenamefont
  {Gehring}, \citenamefont {Auli}, \citenamefont {Grangier}, \citenamefont
  {Yarats},\ and\ \citenamefont {Dauphin}}]{gehring2017convolutional}%
  \BibitemOpen
  \bibfield  {author} {\bibinfo {author} {\bibfnamefont {J.}~\bibnamefont
  {Gehring}}, \bibinfo {author} {\bibfnamefont {M.}~\bibnamefont {Auli}},
  \bibinfo {author} {\bibfnamefont {D.}~\bibnamefont {Grangier}}, \bibinfo
  {author} {\bibfnamefont {D.}~\bibnamefont {Yarats}}, \ and\ \bibinfo {author}
  {\bibfnamefont {Y.~N.}\ \bibnamefont {Dauphin}},\ }\href@noop {} {\bibfield
  {journal} {\bibinfo  {journal} {arXiv:1705.03122}\ } (\bibinfo {year}
  {2017})}\BibitemShut {NoStop}%
\bibitem [{\citenamefont {Zhang}\ and\ \citenamefont
  {Wallace}(2015)}]{zhang2015sensitivity}%
  \BibitemOpen
  \bibfield  {author} {\bibinfo {author} {\bibfnamefont {Y.}~\bibnamefont
  {Zhang}}\ and\ \bibinfo {author} {\bibfnamefont {B.}~\bibnamefont
  {Wallace}},\ }\href@noop {} {\bibfield  {journal} {\bibinfo  {journal}
  {arXiv:1510.03820}\ } (\bibinfo {year} {2015})}\BibitemShut {NoStop}%
\bibitem [{\citenamefont {Kirillov}\ \emph {et~al.}(2015)\citenamefont
  {Kirillov}, \citenamefont {Schlesinger}, \citenamefont {Forkel},
  \citenamefont {Zelenin}, \citenamefont {Zheng}, \citenamefont {Torr},\ and\
  \citenamefont {Rother}}]{kirillov2015generic}%
  \BibitemOpen
  \bibfield  {author} {\bibinfo {author} {\bibfnamefont {A.}~\bibnamefont
  {Kirillov}}, \bibinfo {author} {\bibfnamefont {D.}~\bibnamefont
  {Schlesinger}}, \bibinfo {author} {\bibfnamefont {W.}~\bibnamefont {Forkel}},
  \bibinfo {author} {\bibfnamefont {A.}~\bibnamefont {Zelenin}}, \bibinfo
  {author} {\bibfnamefont {S.}~\bibnamefont {Zheng}}, \bibinfo {author}
  {\bibfnamefont {P.}~\bibnamefont {Torr}}, \ and\ \bibinfo {author}
  {\bibfnamefont {C.}~\bibnamefont {Rother}},\ }\href@noop {} {\bibfield
  {journal} {\bibinfo  {journal} {arXiv:1511.05067}\ } (\bibinfo {year}
  {2015})}\BibitemShut {NoStop}%
\bibitem [{\citenamefont {Song}\ \emph {et~al.}(2019)\citenamefont {Song},
  \citenamefont {Huang},\ and\ \citenamefont {Ruan}}]{song2019abstractive}%
  \BibitemOpen
  \bibfield  {author} {\bibinfo {author} {\bibfnamefont {S.}~\bibnamefont
  {Song}}, \bibinfo {author} {\bibfnamefont {H.}~\bibnamefont {Huang}}, \ and\
  \bibinfo {author} {\bibfnamefont {T.}~\bibnamefont {Ruan}},\ }\href@noop {}
  {\bibfield  {journal} {\bibinfo  {journal} {Multimedia Tools and
  Applications}\ }\textbf {\bibinfo {volume} {78}},\ \bibinfo {pages} {857}
  (\bibinfo {year} {2019})}\BibitemShut {NoStop}%
\bibitem [{\citenamefont {Yu}\ \emph {et~al.}(2018)\citenamefont {Yu},
  \citenamefont {Dohan}, \citenamefont {Luong}, \citenamefont {Zhao},
  \citenamefont {Chen}, \citenamefont {Norouzi},\ and\ \citenamefont
  {Le}}]{yu2018qanet}%
  \BibitemOpen
  \bibfield  {author} {\bibinfo {author} {\bibfnamefont {A.~W.}\ \bibnamefont
  {Yu}}, \bibinfo {author} {\bibfnamefont {D.}~\bibnamefont {Dohan}}, \bibinfo
  {author} {\bibfnamefont {M.-T.}\ \bibnamefont {Luong}}, \bibinfo {author}
  {\bibfnamefont {R.}~\bibnamefont {Zhao}}, \bibinfo {author} {\bibfnamefont
  {K.}~\bibnamefont {Chen}}, \bibinfo {author} {\bibfnamefont {M.}~\bibnamefont
  {Norouzi}}, \ and\ \bibinfo {author} {\bibfnamefont {Q.~V.}\ \bibnamefont
  {Le}},\ }\href@noop {} {\bibfield  {journal} {\bibinfo  {journal}
  {arXiv:1804.09541}\ } (\bibinfo {year} {2018})}\BibitemShut {NoStop}%
\bibitem [{\citenamefont {Cong}\ \emph {et~al.}(2019)\citenamefont {Cong},
  \citenamefont {Choi},\ and\ \citenamefont {Lukin}}]{CongLukin2019}%
  \BibitemOpen
  \bibfield  {author} {\bibinfo {author} {\bibfnamefont {I.}~\bibnamefont
  {Cong}}, \bibinfo {author} {\bibfnamefont {S.}~\bibnamefont {Choi}}, \ and\
  \bibinfo {author} {\bibfnamefont {M.~D.}\ \bibnamefont {Lukin}},\ }\href@noop
  {} {\bibfield  {journal} {\bibinfo  {journal} {Nature Physics}\ }\textbf
  {\bibinfo {volume} {15}},\ \bibinfo {pages} {1273} (\bibinfo {year}
  {2019})}\BibitemShut {NoStop}%
\bibitem [{\citenamefont {Mitarai}\ \emph {et~al.}(2018)\citenamefont
  {Mitarai}, \citenamefont {Negoro}, \citenamefont {Kitagawa},\ and\
  \citenamefont {Fujii}}]{mitarai2018quantum}%
  \BibitemOpen
  \bibfield  {author} {\bibinfo {author} {\bibfnamefont {K.}~\bibnamefont
  {Mitarai}}, \bibinfo {author} {\bibfnamefont {M.}~\bibnamefont {Negoro}},
  \bibinfo {author} {\bibfnamefont {M.}~\bibnamefont {Kitagawa}}, \ and\
  \bibinfo {author} {\bibfnamefont {K.}~\bibnamefont {Fujii}},\ }\href@noop {}
  {\bibfield  {journal} {\bibinfo  {journal} {Physical Review A}\ }\textbf
  {\bibinfo {volume} {98}},\ \bibinfo {pages} {032309} (\bibinfo {year}
  {2018})}\BibitemShut {NoStop}%
\bibitem [{\citenamefont {Rumelhart}\ \emph {et~al.}(1988)\citenamefont
  {Rumelhart}, \citenamefont {Hinton}, \citenamefont {Williams} \emph
  {et~al.}}]{rumelhart1988learning}%
  \BibitemOpen
  \bibfield  {author} {\bibinfo {author} {\bibfnamefont {D.~E.}\ \bibnamefont
  {Rumelhart}}, \bibinfo {author} {\bibfnamefont {G.~E.}\ \bibnamefont
  {Hinton}}, \bibinfo {author} {\bibfnamefont {R.~J.}\ \bibnamefont
  {Williams}},  \emph {et~al.},\ }\href@noop {} {\bibfield  {journal} {\bibinfo
   {journal} {Cognitive modeling}\ }\textbf {\bibinfo {volume} {5}},\ \bibinfo
  {pages} {1} (\bibinfo {year} {1988})}\BibitemShut {NoStop}%
\bibitem [{\citenamefont {Guo}\ and\ \citenamefont
  {Poletti}(2021)}]{GuoPoletti2020a}%
  \BibitemOpen
  \bibfield  {author} {\bibinfo {author} {\bibfnamefont {C.}~\bibnamefont
  {Guo}}\ and\ \bibinfo {author} {\bibfnamefont {D.}~\bibnamefont {Poletti}},\
  }\href@noop {} {\bibfield  {journal} {\bibinfo  {journal} {Physical Review
  E}\ }\textbf {\bibinfo {volume} {103}},\ \bibinfo {pages} {013309} (\bibinfo
  {year} {2021})}\BibitemShut {NoStop}%
\bibitem [{\citenamefont {Kingma}\ and\ \citenamefont
  {Ba}(2014)}]{kingma2014adam}%
  \BibitemOpen
  \bibfield  {author} {\bibinfo {author} {\bibfnamefont {D.~P.}\ \bibnamefont
  {Kingma}}\ and\ \bibinfo {author} {\bibfnamefont {J.}~\bibnamefont {Ba}},\
  }\href@noop {} {\bibfield  {journal} {\bibinfo  {journal} {arXiv:1412.6980}\
  } (\bibinfo {year} {2014})}\BibitemShut {NoStop}%
\bibitem [{\citenamefont {Kerenidis}\ \emph {et~al.}(2019)\citenamefont
  {Kerenidis}, \citenamefont {Landman},\ and\ \citenamefont
  {Prakash}}]{KerenidisPrakash2019}%
  \BibitemOpen
  \bibfield  {author} {\bibinfo {author} {\bibfnamefont {I.}~\bibnamefont
  {Kerenidis}}, \bibinfo {author} {\bibfnamefont {J.}~\bibnamefont {Landman}},
  \ and\ \bibinfo {author} {\bibfnamefont {A.}~\bibnamefont {Prakash}},\
  }\href@noop {} {\bibfield  {journal} {\bibinfo  {journal} {arXiv:1911.01117}\
  } (\bibinfo {year} {2019})}\BibitemShut {NoStop}%
\bibitem [{\citenamefont {Henderson}\ \emph {et~al.}(2020)\citenamefont
  {Henderson}, \citenamefont {Shakya}, \citenamefont {Pradhan},\ and\
  \citenamefont {Cook}}]{henderson2020quanvolutional}%
  \BibitemOpen
  \bibfield  {author} {\bibinfo {author} {\bibfnamefont {M.}~\bibnamefont
  {Henderson}}, \bibinfo {author} {\bibfnamefont {S.}~\bibnamefont {Shakya}},
  \bibinfo {author} {\bibfnamefont {S.}~\bibnamefont {Pradhan}}, \ and\
  \bibinfo {author} {\bibfnamefont {T.}~\bibnamefont {Cook}},\ }\href@noop {}
  {\bibfield  {journal} {\bibinfo  {journal} {Quantum Machine Intelligence}\
  }\textbf {\bibinfo {volume} {2}},\ \bibinfo {pages} {1} (\bibinfo {year}
  {2020})}\BibitemShut {NoStop}%
\bibitem [{VQC()}]{VQC}%
  \BibitemOpen
  \href@noop {} {\bibinfo  {journal} {Available on GitHub at
  https://github.com/guochu/VQC.jl}\ }\BibitemShut {NoStop}%
\bibitem [{\citenamefont {Aleksandrowicz}\ \emph {et~al.}(2019)\citenamefont
  {Aleksandrowicz}, \citenamefont {Alexander}, \citenamefont {Barkoutsos},
  \citenamefont {Bello}, \citenamefont {Ben-Haim}, \citenamefont {Bucher},
  \citenamefont {Cabrera-Hern{\'a}ndez}, \citenamefont {Carballo-Franquis},
  \citenamefont {Chen}, \citenamefont {Chen} \emph {et~al.}}]{qiskit}%
  \BibitemOpen
\bibfield  {journal} {  }\bibfield  {author} {\bibinfo {author} {\bibfnamefont
  {G.}~\bibnamefont {Aleksandrowicz}}, \bibinfo {author} {\bibfnamefont
  {T.}~\bibnamefont {Alexander}}, \bibinfo {author} {\bibfnamefont
  {P.}~\bibnamefont {Barkoutsos}}, \bibinfo {author} {\bibfnamefont
  {L.}~\bibnamefont {Bello}}, \bibinfo {author} {\bibfnamefont
  {Y.}~\bibnamefont {Ben-Haim}}, \bibinfo {author} {\bibfnamefont
  {D.}~\bibnamefont {Bucher}}, \bibinfo {author} {\bibfnamefont {F.~J.}\
  \bibnamefont {Cabrera-Hern{\'a}ndez}}, \bibinfo {author} {\bibfnamefont
  {J.}~\bibnamefont {Carballo-Franquis}}, \bibinfo {author} {\bibfnamefont
  {A.}~\bibnamefont {Chen}}, \bibinfo {author} {\bibfnamefont {C.-F.}\
  \bibnamefont {Chen}},  \emph {et~al.},\ }\href@noop {} {\bibfield  {journal}
  {\bibinfo  {journal} {Accessed on: Mar}\ }\textbf {\bibinfo {volume} {16}}
  (\bibinfo {year} {2019})}\BibitemShut {NoStop}%
\end{thebibliography}%

\end{document}